\documentclass{article}
\usepackage{jheppub}
\usepackage[toc,page]{appendix}
\usepackage{graphicx}
\usepackage{subcaption}
\usepackage{amsmath,amssymb,amscd,amsfonts,mathtools}
\usepackage{slashed}

\hypersetup{colorlinks=true, allcolors=blue}
\usepackage{caption}
\usepackage{xcolor}
\usepackage{hyperref}

\DeclareMathOperator{\Tr}{Tr}
\def\tir{{\Tilde{r}}}

\title{Holographic $a$-functions and Boomerang RG Flows}

\author[1]{Elena C\'aceres,}
\author[1]{Rodrigo Castillo V\'asquez,}
\author[2]{Karl Landsteiner,}
\author[3]{Ignacio Salazar Landea}

\affiliation[1]{Theory Group, Weinberg Institute, Department of Physics, University of Texas, Austin, TX 78712, USA.}
\affiliation[2]{Instituto de F\'\i sica Te\'orica UAM-CSIC, C/ Nicol\'as Cabrera 13-15, Campus Cantoblanco, 28049, Spain }
\affiliation[3]{Instituto de F\'\i sica de La Plata - CONICET, C.C. 67, 1900 La Plata, Argentina}

\emailAdd{elenac@utexas.edu, rcastillov@utexas.edu, karl.landstieiner@csic.es, peznacho@gmail.com}

\begin{document}

\abstract{ We use the radial null energy condition to construct a monotonic $a$-function for a certain type of non-relativistic holographic RG flows. We test our $a$-function in three different geometries that feature a Boomerang RG flow, characterized by a domain wall between two AdS spaces with the same AdS radius, but with different (and sometimes direction-dependent) speeds of light. We find that the $a$-function monotonically decreases and goes to a constant in the asymptotic regimes of the geometry. Using the holographic dictionary in this asymptotic AdS spaces, we find that the $a$-function not only reads the fixed point central charge but also the speed of light, suggesting what the correct RG charge might be for non-relativistic RG flows. } 

\maketitle


\section{Introduction}

To run along the  Renormalization Group (RG) flow, one has to integrate out degrees of freedom. This fact leads to the very intuitive idea that the RG flow is not invertible, as one is losing degrees of freedom. This leads to two related questions. On the one hand, how do we measure the degrees of freedom in a general interacting QFT?
And on the other, how do we prove that the number of degrees of freedom decreases along the RG flow?

This problem is reasonably well understood in the context of relativistic QFT in dimension $2\leq d \leq 4$. Assuming that the RG runs between two different conformal field theories, one can sit at the fixed points and use the enhanced symmetry to recognize potential candidates to measure the degrees of freedom. These are the so-called RG flow charges. Then one may consider the full RG flow and define an $a$-function interpolating between the UV and IR charges\footnote{We note that these are also called $c$-functions.}. A final step consists of explicitly probing the monotonicity of the $a$-function, or at least that it obeys a sum rule ensuring that $C_{UV}\geq C_{IR}$. Note that conformal maps relate observables at the fixed points. Thus,  many possible $a$-functions may be constructed. For instance, the seminal paper starting this program \cite{Zamolodchikov:1986gt} considered the two point function of the stress-energy tensor in the plane. On the other hand, extensions to higher dimensions used the one point function  of the stress-energy tensor on the sphere \cite{Komargodski:2011vj,Komargodski:2011xv} or the universal term in the entanglement entropy \cite{Casini:2004bw,Casini:2012ei,Casini:2017vbe}.

The situation is very different when we perturb the UV CFT with a relevant operator breaking Lorentz invariance. In those cases, several counter examples to a naive extrapolation of the Lorentz invariant results are known \cite{Swingle:2013zla}. This is probably because the correct RG charges for a non-relativistic QFT might not be exactly those counting the degrees of freedom for the relativistic case. A very simple example where this was correctly understood is the case of running defects in the presence of a conformal bulk. This example is particularly illustrative for our purposes for the following reason: if one considers anomaly arguments the naive extension of relativistic arguments, this perfectly works as an $a$-function \cite{Jensen:2015swa,Wang:2021mdq}. On the other hand, the entropic version is slightly more subtle. When the codimension of the defect is greater than one, the universal part of the entanglement entropy of a sphere is no longer the correct RG charge \cite{Jensen:2018rxu,Kobayashi:2018lil}. To properly read the RG charges, one has to compute the relative entropy comparing the QFT of interest with its conformal fixed point, measured at the null cone of a large sphere. Recognizing the universal terms in the expansion of the such quantity leads to the correct RG charges which can be proved to be monotonic under the RG flow \cite{Casini:2022bsu,Casini:2023kyj}. This simple example of RG-flows without full Lorentz symmetry illustrates on the challenges ahead.

In the context of fully relativistic holographic RG flows, a parallel program has been developed \cite{Freedman:1999gp,Myers:2010xs,Myers:2010tj}. In this case, the holographic charges are the effective $AdS$ radius at the UV and IR geometries. Then, the null energy condition ensures the monotonicity of a properly defined $a$-function. Finally, the holographic dictionary connects the RG charges at the fixed points with their general CFT candidates. These examples serve as guiding lines towards identifying the correct RG charges in strongly coupled systems where one has parametric control of the systems along the full RG flow.

The situation is more complicated in a more general set up, where non-relativistic relevant deformations are considered in the context of AdS/CFT. Some previous efforts include results for Lifshitz holography \cite{Hoyos:2010at,Liu:2012wf} and more recently the so called RG flows across the dimensions \cite{GonzalezLezcano:2022mcd}. The aim of this work is to win some intuition in this direction\footnote{ A (probably incomplete) list of interesting developments in this direction should also include \cite{Ghosh:2018qtg,Arav:2018njv,Hoyos:2020zeg,Jokela:2021knd} }. 
To do so, we will elaborate on the concept of holographic $a$-function. Recall that in holography, a black
hole with matter-induced backreaction describes an RG flow from a UV thermal state
towards an IR fixed point associated with the horizon. 
In \cite{Caceres:2022hei,Caceres:2022smh, Caceres:2023zhl}, the authors showed that imposing  the null energy condition is enough to construct a function that is monotonic along the entire flow\footnote{In \cite{Caceres:2022hei,Caceres:2022smh, Caceres:2023zhl} the authors showed that the $a$-function they constructed is also monotonic in  the \emph{trans-IR}, or behind the horizon, part of the flow. In this work, we focus on the standard UV-IR flow and do not concern ourselves with the trans-IR.}. In the present work, our goal is to 
build a holographic $a$-function, $a_T$, for a class of metrics not included in \cite{Caceres:2022hei,Caceres:2022smh, Caceres:2023zhl},  prove that it is monotonic and apply it to 
three different models in the literature that previously reported Boomerang RG flows\footnote{For previous works reporting Boomerang RG flow geometries see \cite{Chesler:2013qla,Donos:2014gya,Donos:2016zpf,Donos:2017ljs}.}. A boomerang RG flow is a very particular non-relativistic RG flow where the IR geometry is an AdS with the same radius as  the UV. Hence in the deep IR we recover Lorentz invariance but the naive RG charges seem to be the same! This makes explicit the necessity to reformulate our understanding of the RG flow for non-relativistic theories. We will find that although the AdS radius is the same for the UV and IR geometries, to recover the same exact geometry one should re-scale the space-time coordinates. Our $a$-function is sensitive to such re-scaling rendering it a good witness of the RG scale. 

The Boomerang RG flows we will consider here will come from three different patterns of symmetry breaking of the Poincar\'e group.
Firstly we will consider \cite{Donos:2017sba} where the Boomerang RG flow comes from a particular explicit breaking of translation invariance. This is done by hiding the spatial dependence in the phase of some scalars that preserve a global symmetry. This useful trick was widely explored as it allows to study holographic momentum dissipation without  solving PDEs \cite{Andrade:2013gsa,Donos:2013eha}. Furthermore, this model also features, in some regime of parameters, some intermediate AdS geometries with a different AdS radius, making them particularly interesting for our purposes.

Secondly, we will study a particular non-relativistic RG flow built in the context of topological semi-metals \cite{Landsteiner:2015lsa,Landsteiner:2015pdh}. This theory, dubbed as the holographic Weyl semimetal, features a topological phase transition between a Dirac phase and a gapped phase, when adjusting the parameter of a rotation breaking relevant deformation. At the gapped phase, the RG flow is pretty standard in the sense that the effective AdS radius is smaller in the IR than the UV. On the other hand, the Dirac phase features a curious Boomerang RG flow, as the theory recovers rotation invariance in the deep IR, flowing to a CFT with seemingly the same central charge than in the UV. Finally, the quantum critical point is characterized by a Lifshitz geometry in the IR. Furthermore, we also know a free fermionic model featuring this same RG patterns, making it a good toy model to explore these interesting topics further \cite{Grushin:2012mt}.

Finally, we will present a model where the Boomerang RG flow appears as the deep infrared limit of a solution that explicitly breaks Lorentz invariance but spontaneously breaks rotation invariance. This solution was presented in the context of topological semimetals and represents the holographic dual of a system with quadratic band touching \cite{Grandi:2021bsp}. At low enough temperatures, the interactions lift the degeneracy and the ground state consists of a nematic phase. The phase transition can be interpreted as a Berry monopole of charge two splitting into two Berry monopoles of charge one \cite{Grandi:2021jkj}. At zero $T$ the dual geometry consists of a Boomerang domain wall that breaks rotation invariance at intermediate scales.

The main lesson we learned from the three examples studied is that although the IR geometries correspond to AdS spaces with the same AdS radius, the actual IR Minkowski metric features a different speed of light. In fact, this speed of light may even depend on the spatial direction we are looking at. The irreversibility of the RG flow can then be reinterpreted as the monotonicity of the light speed along the RG flow, whith the highest speed corresponding to the UV fixed point. This criteria was first sketched in \cite{Hoyos:2010at} in the context of RG flows interpolating between Lifshitz fixed points (see also \cite{Gubser:2009cg}). In the context of Boomerang RG flows, this answer is even more intuitive as light cones are naturally defined in relativistic field theories, such as the ones we find at the fixed points. 
This is a very intuitive answer and we hope it sheds some light on the difficult question of characterizing the irreversibility of the RG in non-relativistic QFTs.

This paper is organized as follows: in Section \ref{Section:GeneralAFunction} we summarize previous findings regarding the holographic $a$-function for anisotropic flows to then introduce the more general case to be used in this paper. We also prove the monotonicity of this new $a$-function. Then, we proceed to apply this $a$-function to the three models mentioned above: Section \ref{Section:Q-lattice} includes the Q-lattice model with RG flows that display intermediate AdS regimes, Section \ref{Section:HWSM} presents the results for Holographic Weyl Semi-metals, and the case of Holographic Flat Bands is shown in Section \ref{Section:HFB}. Finally, Section \ref{Section:Outlook} presents a general discussion of the results and interesting questions regarding this work that remain unanswered.



\section{A More General $a$-function}\label{Section:GeneralAFunction}

In \cite{Caceres:2022hei}, the authors constructed a monotonic $a$-function for anisotropic flows in $(d+1)$ dimensions. In this section, we present a generalization of their results to include a wider class of metrics.

Let us first review the results in  \cite{Caceres:2022hei}. The authors consider backgrounds described by a metric of the form
\begin{equation}\label{eq:DomainWallMetric}
    \mbox{d}s^2 = e^{2A(\rho)} \left[-f(\rho)^2 dt^2+e^{2 \mathcal{X}(\rho)} d\vec{x}_1^2+d\vec{x}_2^2 \right]+d\rho^2~,
\end{equation}
where $t\in \mathbb{R}$, $\rho \geq 0$, $\vec{x}_1 \in \mathbb{R}^{\delta_1}$, $\vec{x}_2 \in \mathbb{R}^{\delta_2}$, and $\delta_1+\delta_2=d-1$. Constant-$\rho$ slices are required to be Lorentzian on $\rho>0$, which implies $f(\rho)>0$. When working with black hole solutions, the function $f(\rho)$ has a simple root at the horizon $\rho=0$. 

The requirement to have an asymptotically AdS space with curvature radius $L$ at large $\rho$ translates into the conditions
\begin{equation}\label{eq:BdryCondDWFuncs}
    A(\rho) \xrightarrow[\rho \rightarrow \infty]{} \frac{\rho}{L}~,~~~~~~~~~~~~~~~\mathcal{X}(\rho) \xrightarrow[\rho \rightarrow \infty]{}0~,~~~~~~~~~~~~~~~f(\rho) \xrightarrow[\rho \rightarrow \infty]{} 1~.
\end{equation}
It was shown in  \cite{Caceres:2022hei} that using the radial null vector
\begin{equation}\label{eq:RadialNull}
    k^{\mu} = \frac{e^{-A(\rho)}}{f(\rho)} \delta^{\mu}_t + \delta^{\mu}_{\rho},
\end{equation}
the radial Null Energy Condition (NEC)  for \eqref{eq:DomainWallMetric} is 
\begin{equation}\label{eq:DW_NECcondition}
    \mathcal{C}(\rho) \frac{d}{d\rho} \left[\mathfrak{a}(\rho)^{1/(d-1)} \right] - \mathcal{K}(\rho)^2 \geq 0~,
\end{equation}
where $\mathcal{C}$ was shown to be a positive-definite real function, $\mathfrak{a}^{1/(d-1)}$ is the principal branch of the (d-1)th root of some real function $\mathfrak{a}$, and $\mathcal{K}$ is some real function. Thus, the  NEC guarantees that  
\begin{equation}\label{eq:Ineq_a}
    \frac{d}{d\rho} \left[\mathfrak{a}(\rho)^{1/(d-1)} \right]  \geq 0~,
\end{equation}
where, 
\begin{eqnarray}\label{eq:DW_IdentificationsRNEC}
\mathfrak{a}(\rho) &= &e^{-\delta_1 \mathcal{X}(\rho)} \left[\frac{(d-1) f(\rho)}{\delta_1 (A'(\rho)+\mathcal{X}'(\rho))+\delta_2 A'(\rho)} \right]^{d-1}~,\nonumber\\
\mathcal{C}(\rho) &= &\frac{1}{(d-1)f(\rho)} \left[\delta_1 (A'(\rho)+\mathcal{X}'(\rho))+\delta_2 A'(\rho) \right]^2 e^{\delta_1 \mathcal{X}(\rho)/(d-1)}~,\\
\mathcal{K}(\rho) &= &\sqrt{\frac{\delta_1 \delta_2}{d-1}} \mathcal{X}'(\rho)~.\nonumber
\end{eqnarray}

Thus the $a$-function is identified as $a_T(\rho) \sim \mathfrak{a}(\rho)$,
\begin{equation}\label{eq:aFuncAnisotropic}
    a_T \sim e^{-\delta_1 \mathcal{X}(\rho)} \left[\frac{(d-1) f(\rho)}{\delta_1 (A'(\rho)+\mathcal{X}'(\rho))+\delta_2 A'(\rho)} \right]^{d-1}~.\\
\end{equation}

The subscript in $a_T$ emphasizes that this is a black hole background. It was shown that in \cite{Caceres:2022hei}  that this $a$-function is monotonic both in the exterior and the interior of the black hole. The interior of the black hole can be accessed by performing an analytic continuation. However, the interior of the black hole does not concern us here since we will deal only with zero-temperature backgrounds. 

\subsection{Generalizing the $a$-function}\label{sec:generalizing_a}
Some Boomerang flows are not described by the class of metrics considered in \cite{Caceres:2022hei}. Here, we derive an $a$-function for backgrounds of the form, 

\begin{equation}\label{eq:GeneralMetric}
    ds^2=e^{2A(\rho)} \left[-f( \rho)^2 dt^2+ dx^2 + dy^2 + 2H(\rho)dx~ dy + \mathcal{Z}(\rho) d\vec{z}^2 \right]+d\rho^2~,
\end{equation}
where $|H(\rho)|<1$ for all values of $\rho$, and $\mathcal{Z}(\rho)$ is a non-negative function. The following boundary conditions guarantee that this spacetime is asymptotically AdS:
\begin{equation}\label{eq:BdryCondXs}
   A(\rho) \xrightarrow[\rho \rightarrow \infty]{} \frac{\rho}{L}\, , \hspace{.5in} \mathcal{Z}(\rho) \xrightarrow[\rho \rightarrow \infty]{} 1\, , \hspace{.5in}  H(\rho) \xrightarrow[\rho \rightarrow \infty]{}  0\, , \hspace{.5in}  f(\rho) \xrightarrow[\rho \rightarrow \infty]{} 1~.
\end{equation}

To derive the $a$-function we proceed as before: impose the NEC using the radial null vector \eqref{eq:RadialNull} and recast the expression obtained in the form \eqref{eq:DW_NECcondition}. We obtain, 
\begin{eqnarray}\label{eq:IdentificationsRNEC}
\mathfrak{a}(\rho) &= &\displaystyle \frac{1}{\left[(1-H(\rho)^2)~ \mathcal{Z}(\rho)^{(d-3)} \right]^{1/2}}~\left[\frac{(d-1) f(\rho)}{(d-1)A'(\rho)+\frac{(d-3)}{2} \frac{\mathcal{Z}'(\rho)}{\mathcal{Z}(\rho)}-\frac{H(\rho) H'(\rho)}{1-H(\rho)^2}} \right]^{d-1}~,\nonumber\\
\mathcal{C}(\rho) &= &\displaystyle \frac{\left[(1-H(\rho)^2)~ \mathcal{Z}(\rho)^{(d-3)} \right]^{1/2(d-1)}}{(d-1)f(\rho)} \left[(d-1)A'(\rho)+\frac{(d-3)}{2} \frac{\mathcal{Z}'(\rho)}{\mathcal{Z}(\rho)}-\frac{H(\rho) H'(\rho)}{1-H(\rho)^2} \right]^2 ~,\\
\mathcal{K}(\rho) &= &\frac{1}{\sqrt{d-1}}\left[\left( \frac{H'(\rho)}{1-H(\rho)^2} \right)^2 + \frac{(d-3)}{4} \left(\frac{H'(\rho)}{(1+H(\rho))}- \frac{\mathcal{Z}'(\rho)}{\mathcal{Z}(\rho)} \right)^2 + \frac{(d-3)}{4} \left(-\frac{H'(\rho)}{(1-H(\rho))}- \frac{\mathcal{Z}'(\rho)}{\mathcal{Z}(\rho)}  \right)^2 \right]^{1/2}~.\nonumber
\end{eqnarray}
Thus, in this case,  the $a$-function is,
 \begin{equation} \label{eq:a_Boomerang}
 a(\rho) =\displaystyle \frac{1}{\left[(1-H(\rho)^2)~ \mathcal{Z}(\rho)^{(d-3)} \right]^{1/2}}~\left[\frac{(d-1) f(\rho)}{(d-1)A'(\rho)+\frac{(d-3)}{2} \frac{\mathcal{Z}'(\rho)}{\mathcal{Z}(\rho)}-\frac{H(\rho) H'(\rho)}{1-H(\rho)^2}} \right]^{d-1}~
 \end{equation}
 
\subsection{Monotonicity}

To prove that \eqref{eq:a_Boomerang} is a monotonic function, we need to prove that its derivative is positive for all values of $\rho$. To do this, we start off with the identification $a(\rho) = \mathfrak{a}(\rho)$ and then rewrite the derivative of this function as follows:
\begin{equation}
    \frac{d}{d\rho}[\mathfrak{a}(\rho)] = \frac{d}{d\rho} \left[\left(\mathfrak{a}(\rho)^{1/(d-1)} \right)^{(d-1)} \right] =(d-1)\mathfrak{a}(\rho)^{(d-2)/(d-1)} \frac{d}{d\rho} \left[\mathfrak{a}(\rho)^{1/(d-1)} \right]~.
\end{equation}
From \eqref{eq:Ineq_a}, we know the NEC guarantees that the derivative of $\mathfrak{a}(\rho)^{1/(d-1)}$ is non-negative. Therefore, our goal is to prove that 
\begin{equation*}
    \mathfrak{a}(\rho)^{(d-2)/(d-1)} = \displaystyle \frac{1}{\left[(1-H(\rho)^2)~ \mathcal{Z}(\rho)^{(d-3)} \right]^{(d-2)/(2d-2)}}~\left[\frac{(d-1) f(\rho)}{(d-1)A'(\rho)+\frac{(d-3)}{2} \frac{\mathcal{Z}'(\rho)}{\mathcal{Z}(\rho)}-\frac{H(\rho) H'(\rho)}{1-H(\rho)^2}} \right]^{d-2} \geq 0~.
\end{equation*}

The factor $(d-1)f$ is positive for all values of $\rho$. The conditions $|H(\rho)|<1$ and $\mathcal{Z}(\rho) > 0$ guarantee that 
$$
\left[(1-H(\rho)^2)~ \mathcal{Z}(\rho)^{(d-3)} \right]^{(d-2)/(2d-2)} > 0~,
$$
for any $d \geq 3$. What remains is to show that
$$
\left[ (d-1)A'(\rho)+\frac{(d-3)}{2} \frac{\mathcal{Z}'(\rho)}{\mathcal{Z}(\rho)}-\frac{H(\rho) H'(\rho)}{1-H(\rho)^2} \right]^d \geq 0~,
$$
which is automatically true for even $d$. However, we ultimately want to prove that 
\begin{equation}\label{eq:Ineq_Proof}
(d-1)A'(\rho)+\frac{(d-3)}{2} \frac{\mathcal{Z}'(\rho)}{\mathcal{Z}(\rho)}-\frac{H(\rho) H'(\rho)}{1-H(\rho)^2}  \geq 0~,
\end{equation}
for any $d$, be it odd or even. 

To prove \eqref{eq:Ineq_Proof}, we proceed with a proof by contradiction. The first step of such proof consists of assuming that there exists a value of $\rho$, denoted by $\rho_*>0$, for which the left-hand side of \eqref{eq:Ineq_Proof} crosses zero. For $\rho>\rho_*$ the l.h.s. will be positive, which we know already since the conditions \eqref{eq:BdryCondXs} indicate that the expression asymptotes to
$$
(d-1)A'(\rho)+\frac{(d-3)}{2} \frac{\mathcal{Z}'(\rho)}{\mathcal{Z}(\rho)}-\frac{H(\rho) H'(\rho)}{1-H(\rho)^2}~ \xrightarrow[\rho \rightarrow \infty]{}~ \frac{(d-1)}{L} > 0~.
$$
This means we are assuming that the l.h.s. of the previous expression will be negative for $\rho < \rho_*$, zero at $\rho=\rho_*$ and non-negative for $\rho=\rho_*$.

The next step is to take advantage of the analyticity of the functions to perform a Taylor-expansion about this point
\begin{equation} \label{eq:Expansion1}
(d-1)A'(\rho)+\frac{(d-3)}{2} \frac{\mathcal{Z}'(\rho)}{\mathcal{Z}(\rho)}-\frac{H(\rho) H'(\rho)}{1-H(\rho)^2} = k (\rho-\rho_*)^m + \mathcal{O}\left[(\rho-\rho_*)^{m+1} \right]~,
\end{equation}
where the constant $k \neq 0$ and $m \geq 1$ is the degree of the root. Let us evaluate the previous expansion at $\rho=\rho_*+\epsilon$, where $\epsilon > 0$ is an arbitrarily small number, and approximate it by
\begin{equation} \label{eq:Approx1}
    (d-1)A'(\rho_*+\epsilon)+\frac{(d-3)}{2} \frac{\mathcal{Z}'(\rho_*+\epsilon)}{\mathcal{Z}(\rho_*+\epsilon)}-\frac{H(\rho_*+\epsilon) H'(\rho_*+\epsilon)}{1-H(\rho_*+\epsilon)^2} \approx k \epsilon^m~.
\end{equation}
Notice that our assumptions imply $k\epsilon^m>0$. Furthermore, this implies that $k>0$. By taking the derivative with respect to $\rho$ of \eqref{eq:Expansion1} and evaluating it at $\rho=\rho_*+\epsilon$, we obtain
\begin{eqnarray} \label{eq:Approx2}
    (d-1) A''(\rho_*+\epsilon) +\frac{(d-3)}{2} \frac{Z(\rho_*+\epsilon)Z''(\rho_*+\epsilon)-Z'(\rho_*+\epsilon)^2}{Z(\rho_*+\epsilon)^2} \nonumber \\
    -\frac{(1-H(\rho_*+\epsilon)^2) H(\rho_*+\epsilon) H''(\rho_*+\epsilon)+(1+H(\rho_*+\epsilon)^2)H'(\rho_*+\epsilon)^2}{(1-H(\rho_*+\epsilon)^2)^2} &\approx &k m \epsilon^{m-1}~.
\end{eqnarray}
We use \eqref{eq:Approx1} and \eqref{eq:Approx2} to solve for $A'$ and $A''$. So, at $\rho=\rho_*+\epsilon$, we approximately have (dropping the arguments of the functions)
\begin{eqnarray} \label{eq:ApproxA}
    A' &\approx &k \epsilon^m + \frac{H H'}{1-H^2}~,\nonumber \\
    A'' &\approx &k m \epsilon^{m-1} + \frac{(1-H^2) H H''+(1+H^2)H'^2}{(1-H^2)^2} - \frac{(d-3)}{2} \frac{Z Z''-Z'^2}{Z^2} 
\end{eqnarray}

Next we write out the NEC, denoted by $\mathcal{N}(\rho)$, obtained from using the radial null vector \eqref{eq:RadialNull}:
\begin{eqnarray}\label{eq:ExplicitNEC}
    \mathcal{N}(\rho) &= &(d-1) \left[\frac{f'(\rho) A'(\rho)-f(\rho) A''(\rho)}{f(\rho)} \right] +\left[\frac{f'(\rho)}{f(\rho)}- A'(\rho) \right] \left[\frac{(d-3)}{2} \frac{Z'(\rho)}{Z(\rho)} - \frac{H(\rho) H'(\rho)}{1-H(\rho)^2} \right] \nonumber \\ 
    & &- \left[\frac{(d-3)}{2} \frac{Z(\rho)Z''(\rho)-Z'(\rho)^2}{Z(\rho)^2} -\frac{(1-H(\rho)^2) H(\rho) H''(\rho)+(1+H(\rho)^2)H'(\rho)^2}{(1-H(\rho)^2)^2} \right] \nonumber \\
    & &-\left[\frac{(d-3)}{2} \frac{Z'(\rho)^2}{Z(\rho)^2} - \frac{(1+H(\rho)^2)H'(\rho)^2}{2(1-H(\rho)^2)^2} \right]~\geq 0~.
\end{eqnarray}
We will analyze the behavior of $\mathcal{N}(\rho)$ at $\rho=\rho_*+\epsilon$. For this, we plug the approximations \eqref{eq:ApproxA} into the radial NEC \eqref{eq:ExplicitNEC} and group the expression by powers of $\epsilon$. The approximation has three parts: a term proportional to $\epsilon^m$, another part proportional to $\epsilon^{m-1}$, and a third part not multiplied by a power of $\epsilon$. For small $\epsilon$, the expansion is dominated by the $\epsilon^{m-1}$ term so we may truncate the $\mathcal{O}(\epsilon^m)$ part. We are left with
\begin{eqnarray} \label{eq:NECApprox}
    \mathcal{N}(\rho_*+\epsilon) &\approx &-k m \epsilon^{m-1} -\frac{1}{d-1} \left[\left( \frac{H'(\rho_*+\epsilon)}{1-H(\rho_*+\epsilon)^2} \right)^2 + \frac{(d-3)}{4} \left(\frac{H'(\rho_*+\epsilon)}{(1+H(\rho_*+\epsilon))}- \frac{\mathcal{Z}'(\rho_*+\epsilon)}{\mathcal{Z}(\rho_*+\epsilon)} \right)^2 \right. \nonumber \\
    & &\left. + \frac{(d-3)}{4} \left(-\frac{H'(\rho_*+\epsilon)}{(1-H(\rho_*+\epsilon))}- \frac{\mathcal{Z}'(\rho_*+\epsilon)}{\mathcal{Z}(\rho_*+\epsilon)}  \right)^2 \right] \geq 0~.
\end{eqnarray}
Since $k~ \epsilon^m >0$, $\epsilon>0$ and $m \geq 1$, it is clear that the right-hand side of \eqref{eq:NECApprox} is negative. In fact, the second term on the right-hand side of the approximation is equal to $\mathcal{K}^2$. This represents a violation of the NEC, creating a contradiction. Therefore, we know that \eqref{eq:Ineq_Proof} must be true for all $\rho>0$.

\section{Boomerang Flows with Intermediate Conformal Regimes} \label{Section:Q-lattice}

In \cite{Donos:2017sba}, a bottom-up model that flows from a UV AdS$_5$ fixed point to the same AdS$_5$ in the IR was constructed. It was also found that, for large enough values of the tunable parameter in the model, it is possible to have an AdS$_5$ scaling regime for intermediate values of the radial coordinate. We would like to know how all these AdS fixed points are captured by the $a$-function, in the sense that our construction should be able to discriminate one AdS fixed point from the others due to its monotonic behavior. 

Holographic Q-lattice constructions \cite{Donos:2013eha} allow for an ansatz of the bulk fields whose dependence on the spatial coordinates of the CFT is solved exactly. This realization is possible by taking advantage of a global symmetry of the bulk spacetime. Such constructions conveniently lead to a set of ordinary differential equations for a set of fields and metric functions, which depend only on the bulk radial coordinate. The action of the $(4+1)$-dimensional Q-lattice model that we are going to consider is 
\begin{equation}\label{eq:QLat_Action}
S =\frac{1}{2 \kappa^2}\int d^5x\,\sqrt{-g}\Big(R+12+\mathcal{L}_z \Big)~,
\end{equation}
where $\mathcal{L}_z$ describes a sigma model for three complex scalars $z^\alpha$ and has a $U(1)^3$ global symmetry. This can be achieved by considering
\begin{equation}\label{eq:QLat_Lz}
\mathcal{L}_z = \sum_\alpha\left(-\frac{1}{2}\,\partial_\mu z^{\alpha}\partial^\mu \bar{z}^{\bar{\alpha}}
-\frac{1}{2}m^2\,z^{\alpha}\bar{z}^{\bar{\alpha}}-\frac{1}{3}\xi\,\left(z^{\alpha}\bar{z}^{\bar{\alpha}}\right)^2\right)\,,
\end{equation}
where $m^2,~\xi$ are two free parameters. The equations of motion admit an $AdS_5^0$ vacuum solution with $z^\alpha=0$ 
which is dual to a CFT in $d=4$. For the rest of this section, we set $2 \kappa^2=L=1$.

The ansatz for the bulk metric is given by
\begin{align}\label{eq:QLat_Metric}
 ds^2 &= -g(r)e^{-\chi(r)} dt^2+\frac{ dr^2}{g(r)}+r^2dx^\alpha dx^\alpha
 ~,\nonumber \\
z^{\alpha} &= \gamma(r)e^{i k\, x^\alpha}~,
\end{align}
where $x^\alpha \in \{x,y,z\}$ are the spatial directions of the field theory and $k$ is the characteristic wave number of the spatial deformation. The UV boundary is located at $r\rightarrow \infty$. By construction, this ansatz is invariant under a simultaneous translation and a $U(1)^3$ transformation. 

The boomerang RG flows will go from a UV fixed point described by AdS$_5^0$ to the same AdS$_5^0$ in the IR. This AdS solution is characterized by $L^2_0=1$ and $\gamma_0=0$. In order to have deformations by relevant operators that break translation invariance in the dual CFT, the scaling dimension must satisfy $\Delta_0=2+\sqrt{4+m^2}<4$.

By choosing the right values for the fixed parameters $m^2$ and $\xi$ it is also possible to have an intermediate regime described by another AdS solution, which we call AdS$_5^c$. This solution is characterized by 
\begin{equation}
    L_c^2=\frac{64 \xi}{3 m^2 + 64 \xi}~,~~~~~~~~~~\gamma_c=\pm \sqrt{\frac{-3 m^2}{4 \xi}}~.
\end{equation}
The scaling dimension must satisfy $\Delta_c=2+\sqrt{4-2 m^2 L_c^2}>4$ in order to have an irrelevant scalar operator in the dual CFT. 

Following what is done in \cite{Donos:2017sba}, the choice of values $m^2=-15/4$, $\xi=675/512$ translates into $L_c^2=2/3$, $\gamma_c=\pm \sqrt{32/15}$ and the scaling dimensions $\Delta_0=5/2$, $\Delta_c=5$. We work with the positive root of $\gamma_c$.

\subsection{Constructing the Numerical Solution}

We will construct a solution for the Q-lattice model that displays both boomerang behavior and an intermediate AdS scaling. For this, we start with the equations of motion for the field and metric functions in \eqref{eq:QLat_Metric}
\begin{eqnarray}\label{eq:QLat_EOM}
    0 &= &\chi'(r) + r \gamma'(r)^2~, \nonumber\\
    0 &= &g'(r)+g(r) \left(\frac{1}{2}r \gamma'(r)^2+\frac{2}{r} \right) + \gamma(r)^2 \frac{(k^2+m^2 r^2)}{2r} + \frac{1}{3} \xi r \gamma(r)^4-4 r~,\nonumber\\
    0 &= &\gamma''(r)+\gamma'(r) \left(\frac{g'(r)}{g(r)} - \frac{\chi'(r)}{2} + \frac{3}{r} \right) - \gamma(r) \frac{(k^2+m^2 r^2)}{r^2 g(r)}-\frac{4\xi \gamma(r)^3}{3 g(r)}~,
\end{eqnarray}
where the prime denotes the derivative with respect to $r$.

\begin{figure}[h]
    \centering
    \includegraphics[width=0.5\textwidth]{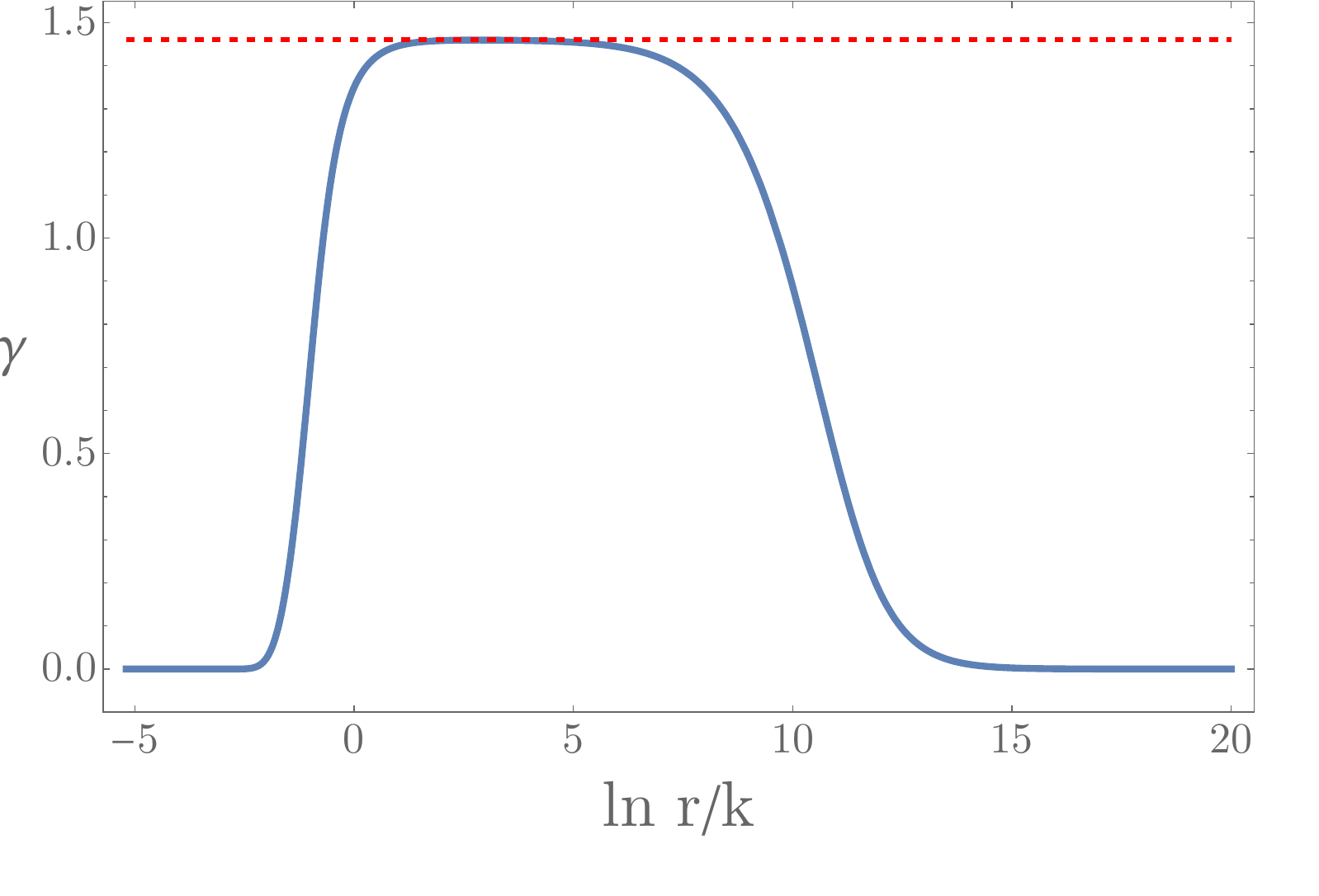}\hfill \includegraphics[width=0.5\textwidth]{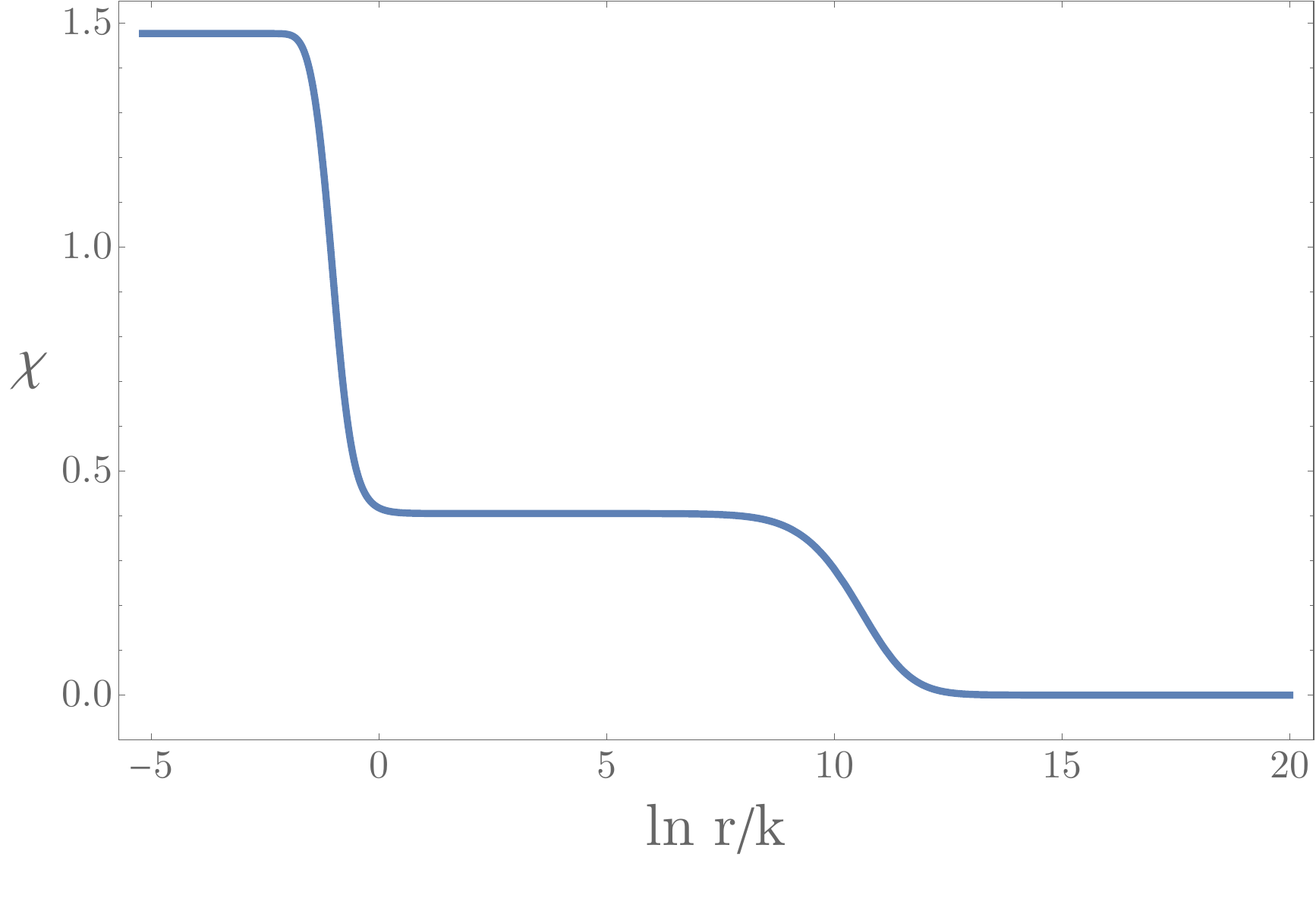}\hfill \includegraphics[width=0.5\textwidth]{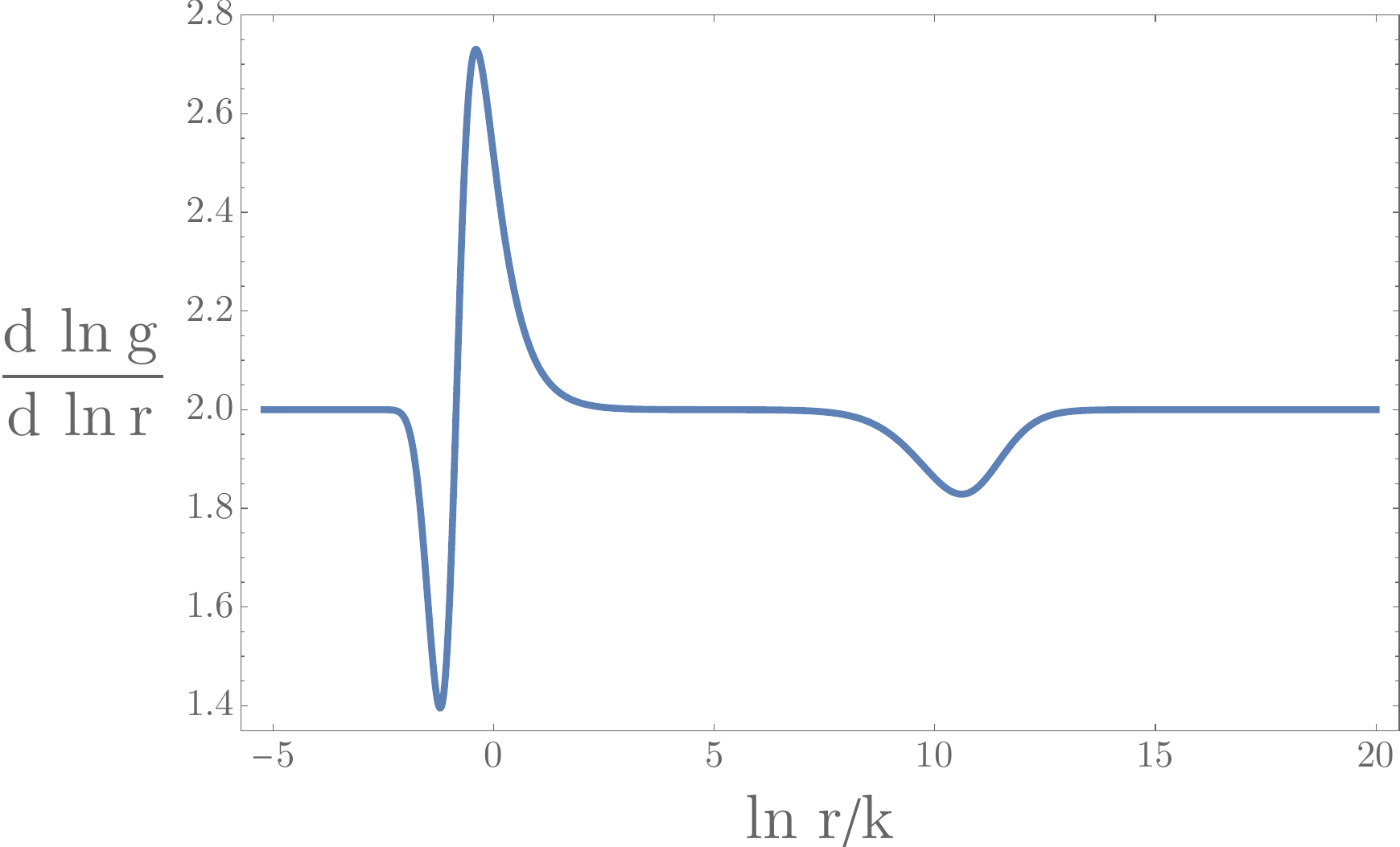}
    \caption{The field and metric functions for $\Gamma/k^{3/2} =10^7$. The function $\gamma(r)$ takes the value $\gamma_c=\sqrt{32/15}$ (red, dashed) in the intermediate AdS regime, as expected.\\}
    \label{fig:MetricFnIntermediate}
\end{figure}

We numerically construct our solutions by integrating the equations of motion from the IR towards the UV. Expanding around the deep IR ($r \rightarrow 0$)
\begin{eqnarray}\label{eq:QLat_IRExpansion}
    g(r) &= &r^2\left[1-\frac{k^3}{4r^3} e^{-2k/r} \left(-3+2\frac{k}{r} \right) \left(\frac{C_\gamma}{k^{3/2}} \right)^2 + \cdots  \right]~,\nonumber\\
    \chi(r) &= &\chi_0 - e^{-2k/r} \frac{1}{16} \left(3+6\frac{k}{r}+6\frac{k^2}{r^2}-8\frac{k^3}{r^3} + 8\frac{k^4}{r^4} \right) \left(\frac{C_\gamma}{k^{3/2}} \right)^2 + \cdots ~,\nonumber \\
    \gamma(r) &= &\frac{k^{3/2}}{r^{3/2}} e^{-k/r} \left(\frac{C_\gamma}{k^{3/2}} \right) + \cdots~, 
\end{eqnarray}
where $k$ is the wave number that parametrizes the spatial deformation. $C_\gamma$ and $\chi_0$ are 
 integration constants that we will use to shoot towards the desired  UV ($r \rightarrow \infty$)
\begin{eqnarray}
    g(r) &= &r^2 + \cdots~,\nonumber\\
    \chi(r) &= &\chi_{UV} + \cdots~,\nonumber\\
    \gamma(r) &= &\Gamma r^{\Delta_0-4} + \cdots~.
\end{eqnarray}
Here $\Gamma$ parametrizes the  relevant deformation in the UV CFT, while choosing the UV speed of light to be one implies $\chi_{UV}=0$. We will organize the different solutions for this Q-lattice model by the dimensionless number $\Gamma/k^{3/2}$. Among these solutions, we will be particularly interested in solutions that not only feature a Boomerang RG flow, but also an intermediate scaling regime with a different AdS radius. This is the case of the solutions that we show in Figure \ref{fig:MetricFnIntermediate} by fixing $\Gamma/k^{3/2}=10^7$.

\subsection{Capturing the Different Regimes with the $a$-function}

The motivation to analyze this Q-lattice model is two-fold. First, we want to determine how the $a$-function captures the boomerang behavior of the numerical solution that we constructed. The second motivation is to explore how such $a$-function captures the intermediate AdS regime in this same solution. Since the $a$-function is monotonic, we expect that it will take different values at each fixed point, thus making it a valuable tool to discriminate between different AdS regimes even in boomerang cases, where the UV and IR are described by the same theory.

To obtain the $a$-function for this theory, we must rewrite \eqref{eq:QLat_Metric} as
\begin{equation}\label{eq:QLat_Metric2}
    ds^2 = r^2 \left[-\frac{g(r) e^{-\chi(r)}}{r^2} dt^2 + dx^2 + dy^2 +dz^2 \right] + \frac{dr^2}{g(r)}~.
\end{equation}
By performing the change of coordinates $r \rightarrow \rho$, it is possible to make the following identifications
\begin{eqnarray}\label{eq:QLat_CoordChange}
    &\displaystyle e^{2A(\rho)}=r^2~,~~~~~~~~~~f(\rho)^2=\frac{g(r) e^{-\chi(r)}}{r^2}~,~~~~~~~~~&\displaystyle d\rho = \frac{dr}{\sqrt{g(r)}}~,\nonumber \\
    &\mathcal{Z}(\rho)=1~~,~~~~~~~&H(\rho)=0~.
\end{eqnarray}
This allows us to rewrite the metric \eqref{eq:QLat_Metric2} in the domain-wall form
\begin{equation}
      ds^2 = e^{2A(\rho)} \left[-f(\rho)^2 dt^2 + dx^2 + dy^2 + dz^2 \right] + d\rho^2~.
\end{equation}
It can be shown that this identification satisfies the boundary conditions \eqref{eq:BdryCondXs}. With this, it is simple to find the $a$-function of this boomerang model by using \eqref{eq:IdentificationsRNEC} $d-1 = 3$. To convert back to the $r$ coordinate, we need the relations
\begin{equation}
    \frac{dA(\rho)}{d\rho}=\frac{\sqrt{g(r)}}{r}~,~~~~~~~~~~f(\rho) = \frac{\sqrt{g(r)}~ e^{-\chi(r)/2}}{r}~.
\end{equation}
So, the $a$-function in terms of the $r$ coordinate is given by
\begin{equation}
   \displaystyle a(r) = e^{-3 \chi(r)/2}~.
\end{equation}
Its derivative with respect to the energy scale $\rho$ is
\begin{equation}
    \frac{da}{d\rho}(r) = -\frac{3}{2} \sqrt{g(r)}~ \chi'(r)~ e^{-3 \chi(r)/2}~,
\end{equation}
where the prime denotes the derivative with respect to $r$. We notice that in this first example the expression for the $a$-function is particularly simple and was studied as a proxy of the RG irreversibility in some seminal papers as \cite{Donos:2017sba,Gauntlett:2018vhk,Gubser:2009gp}, as the negativity of $\chi'$ is immediate from the equations of motion. In the following sections we will show further examples where the constraints coming from the radial NEC are not obvious at first sight. Furthermore, it seems that the radial NEC will be a good guiding principle giving sensible answers in more general situations.   \\

\begin{figure}[h]
    \centering
    \includegraphics[width=0.6\textwidth]{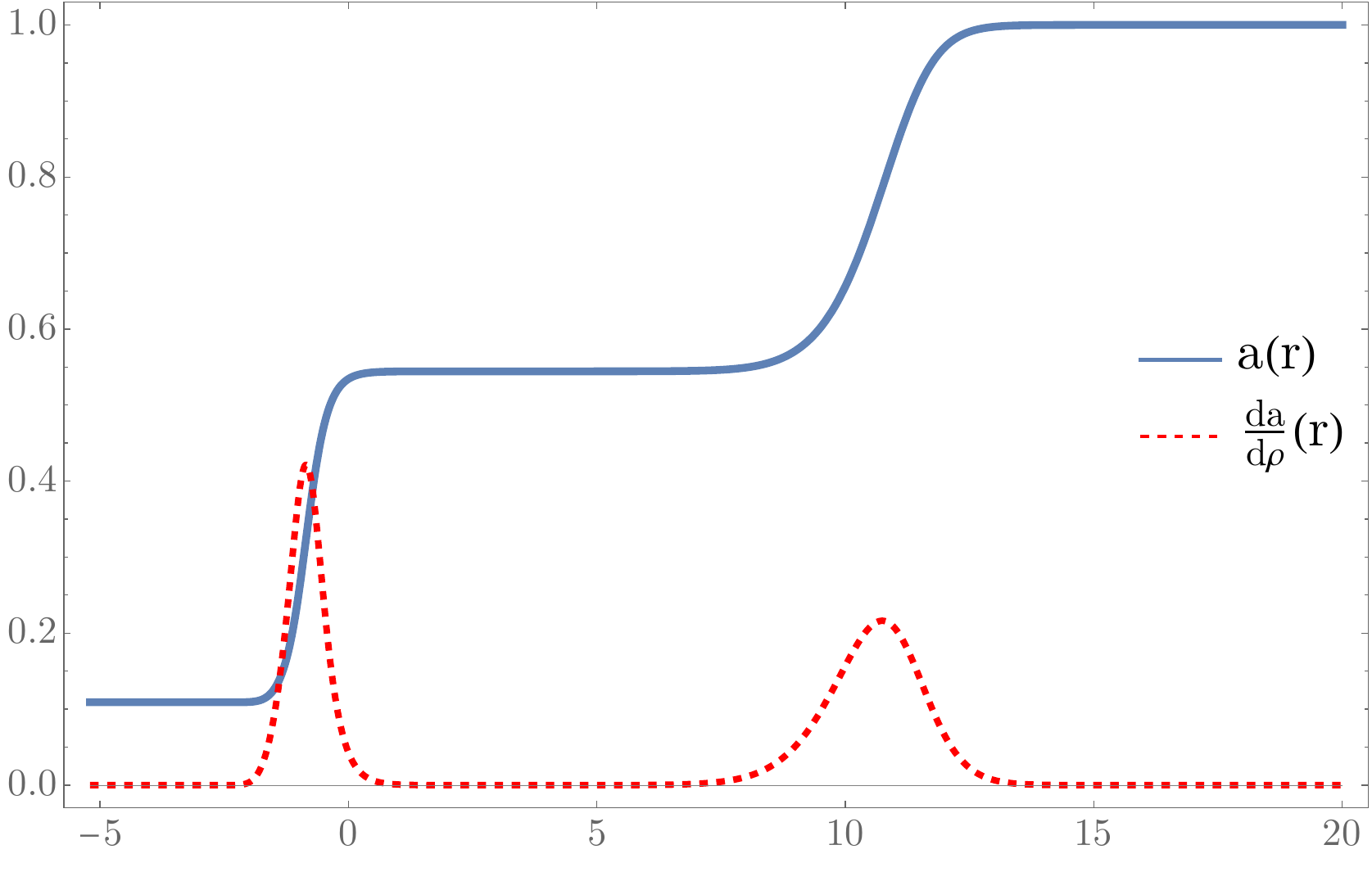}
    \caption{The $a$-function (blue, solid line) and its derivative $da/d\rho$ (red, dashed) for $\Gamma/k^{3/2}=10^7$.}
    \label{fig:aFunctionIntermediate}
\end{figure}

We now evaluate $a(r)$ and $\displaystyle \frac{da}{d\rho}(r)$ in our numerical solutions and show the corresponding result in Figure \ref{fig:aFunctionIntermediate}. The $a$-function captures the three different AdS fixed points very well, as it becomes constant for both UV and IR AdS$_5^0$ and for the intermediate AdS$_5^c$. Furthermore, the $a$-function can differentiate each AdS as it takes different values in each, thanks to its monotonicity. It is worth noting that the derivative with respect to the energy scale $\rho$ becomes zero for each of these regimes, too. It is important to point out that, even though it is the bulk coordinate $\rho$ what is identified with the energy scale that parametrizes the RG flow \cite{Balasubramanian:1999jd}, we present the $a$-function and its derivative with respect to the energy scale in terms of the coordinate $r$. This choice is the more convenient one since the numerical solution for \eqref{eq:QLat_EOM} was found in terms of $r$, and it can be checked by numerical integration of \eqref{eq:QLat_CoordChange} that $r$ and $\rho$ are directly related by a scaling factor only. This means that, even if we show the $a$-function and its derivative in terms of $r$, the behavior displayed by the plots in Figure \ref{fig:aFunctionIntermediate} will have a direct correspondence with what we would observe if we obtained the plots using the $\rho$ coordinate. 

The $a$-function for this metric is particularly simple and it can be re-stated as the monotonicity of the lapse function as we dive into the IR geometry. This reflects the fact that in order to recover the exact same $AdS$ one should make some re-scalings of the time and hence a sensitive irreversible $a$-function should take that into account. In this particular system the irreversibility of the RG flow can be re-stated in a very simple and physical way: The UV speed of light should be the larger than the IR (or any intermediate) speed of light (times the corresponding AdS radius). To make this statement more precise, we can explicitly replace the a function in the AdS geometries by their corresponding speed of light $c$ and AdS radius $L$
\begin{equation}
    a(r)= e^{-3/2\xi}\to c^3 L^3
    \label{afnads}
\end{equation}
Away from the fixed points, the geometry is no longer AdS and the $a$-function does not have a clear interpretation, but interpolates monotonically between the corresponding fixed point values.

It is worth now comparing with the entropic $a$-function defined in \cite{Donos:2017sba}. Such function is defined directly from the entanglement entropy of a strip and its monotonicity can be proved for relativistic holographic RG flows \cite{Myers:2012ed}\footnote{Notice that the known proof for general QFT is defined in terms of a spherical entangling surface \cite{Casini:2017vbe}.}. Such $a$-function properly reads the corresponding AdS radius but has no information about the speed of light. Hence it recognizes when the geometry approaches an AdS region but it is not monotonic along the RG flow. In contrast, our $a$-function is completely monotonic, captures the AdS regimes and even tells them apart from each other. It would be an important step forward to find a sensible entropic measure capable of reading the corresponding speed of light, but such development is beyond the scope of this article.


\section{Holographic Weyl Semimetals} \label{Section:HWSM}
Weyl semi-metals are a relatively new class of materials whose electronics can be described by an effective Dirac equation \cite{RevModPhys.90.015001}. As such they are similar to graphene with the key difference that Weyl semi-metals are three-dimensional materials. Therefore the effective Dirac spinor describing the electronic wave function can be decomposed into left- and right-chiral Weyl spinors. It turns out that these two components live at different points in the Brillouin zone of the material (momentum space).
The Nilsen-Ninomiya theorem guarantees that left- and right-handed fermions always come in pairs. A field theoretical model to describe Weyl semi-metals has been introduced in \cite{Grushin:2012mt} and is given by the Dirac like equation
\begin{equation}
    \left( i \slashed \partial - \gamma_5 \vec\gamma .\vec b - M\right)\psi =0\,.
\end{equation}
From the high energy point of view $b$ is a constant axial gauge field background. Because of the presence of the mass term one might naively expect that the theory is gapped in the infrared. The infrared physics is however governed by an interplay between the two parameters $b$ and $M$.
For $|b|<|M|$ the theory is indeed gapped in the infrared whereas in the opposite regime $|b|>|M|$ the theory essentially flows to the theory of two Weyl fermions separated in momentum space by a distance $\sqrt{|b|-|M|}$.
In this regime the theory has a non-vanishing anomaly induced Hall conductivity proportional to the distance between the Weyl points in momentum space.

A holographic model for topological Weyl semimetals was first introduced in \cite{Landsteiner:2015lsa} in the probe limit. In \cite{Landsteiner:2015pdh} a simple extension of the model showed that it undergoes a quantum phase transition from a topological nontrivial semimetal with non-vanishing anomalous Hall conductivity to a trivial phase. Further properties of these holographic Weyl semimetals (HWSM) were found and analyzed, such as its odd viscosity in the quantum critical region \cite{Landsteiner_2016:OddV} and the parameter space of its quantum phase transition \cite{Copetti_2017} and the Butterfly velocity across the quantum phase transition \cite{Baggioli:2018afg}.

The holographic model is realized with the $(4+1)$-dimensional action
\begin{align}\
  S=&\int d^5x \sqrt{-g}\bigg[\frac{1}{2\kappa^2}\Big(R+\frac{12}{L^2}\Big)-\frac{1}{4}F^2-\frac{1}{4}F_5^2  
+\frac{\alpha}{3}\epsilon^{\mu\nu\rho\sigma\tau}A_\mu \Big(F^5_{\nu\rho} F^5_{\sigma\tau}+3 F_{\nu\rho} F_{\sigma\tau}\Big)\nonumber\\
&-(D_\mu\Phi)^*(D^\mu\Phi)-V(\Phi)\bigg]\,,\label{eq:HWSM_Action}
\end{align}
 where $\kappa^2$ is the Newton constant, $L$ is the AdS radius and $\alpha$ is the Chern-Simons coupling constant. The electromagnetic $U(1)$ symmetry is represented by the AdS bulk gauge field $V_\mu$ with field strength $F=dV$, while the axial $U(1)$ symmetry is represented by the gauge field $A_\mu$ with field strength $F_5 = dA$. We set $2\kappa^2=L=1$.

The bulk scalar field $\Phi$ has a quartic potential $V(\Phi)=m^2 |\Phi|^2 + \frac{\lambda}{2} |\Phi|^4$. The AdS bulk mass $m^2 L^2=-3$ is chosen such that the scaling dimension of the operator dual to $\Phi$ is three and the dimension of its source is one, which means it can be regarded as the boundary mass $M$. The charge $q$ modulates the mixing between the operators dual to $\Phi$ and $A$ via the covariant derivative.

The boundary mass and the time-reversal and rotation breaking parameter $b$ are introduced by means of the boundary conditions
\begin{equation}\label{eq:HWSM_bcs}
 \lim_{r\rightarrow \infty}\,r\Phi = M~,~~~~~~~~\lim_{r\rightarrow \infty}A_z = b\,.
\end{equation}
\ \\

\subsection{The $a$-function of Holographic Weyl Semimetals}
Let us proceed to calculate the $a$-function for the 
zero-temperature ansatz, 
\begin{eqnarray}\label{eq:HWSM_Metric1}
     ds^2 &= u(r)[-dt^2+dx^2+dy^2] +  \displaystyle  \frac{dr^2}{u(r)} + h(r) dz^2~, \\
     \displaystyle   &A = A_z(r)~dz~~~~~~~~~~~~\Phi = \phi(r)~. \nonumber
\end{eqnarray}
Note that $r \rightarrow \infty$ corresponds to the AdS boundary. The only tunable parameter in this case is $M/b$, and depending on its value we can find two different phases separated by a critical point:
\begin{itemize}
    \item A topological nontrivial phase for $M/b < 0.744$. Integrating of the equations of motion shows that a boomerang flow is obtained from an IR AdS$_5$ to a UV AdS$_5$.
    \item A topological trivial phase for $M/b>0.744$. 
    In this regime the Lorentz preserving parameter $m$ dominates the physics of the system, gaping degrees of freedom. Although Lorentz breaking at intermediates scales, the system flows to an AdS$_5$ in the deep IR, with a smaller AdS radius, in agreement with the naive expectations for RG flows.
    \item The critical point $M/b \simeq 0.744$, described by a Lifshitz-type exact scaling solution in the IR. For the right choice of parameters, the solution corresponds to an asymptotic AdS in the UV. 
    \end{itemize}

We proceed to find the $a$-function for the zero-temperature HWSM. We start by rewriting \eqref{eq:HWSM_Metric1} as
\begin{equation}\label{eq:HWSM_Metric2}
    ds^2 = u(r) \left(-dt^2+dx^2+dy^2+\frac{h(r)}{u(r)}dz^2 \right)+\frac{dr^2}{u(r)}~.
\end{equation}
Then, transforming from $r \rightarrow \rho$ involves the following relations
\begin{eqnarray}\label{eq:HWSM_CoordChange}
    &\displaystyle e^{2A(\rho)}=u(r)~,~~~~~~~\mathcal{Z}(\rho)=\frac{h(r)}{u(r)}~,~~~~~~~ \displaystyle d\rho = \frac{dr}{\sqrt{u(r)}}~,~~~~~~~f(\rho)=1~,~~~~~~~H(\rho)=0~.
\end{eqnarray}
By performing such coordinate transformation, we can recast \eqref{eq:HWSM_Metric2} as
\begin{equation}
    ds^2 = e^{2A(\rho)} \left(-dt^2 + dx^2 + dy^2 + \mathcal{Z}(\rho) dz^2 \right) + d\rho^2~.
\end{equation}
With the metric rewritten in this way, it is straightforward to use \eqref{eq:a_Boomerang} and find the $a$-function. Finally, we transform back from $\rho$ to the $r$ coordinate by means of \eqref{eq:HWSM_CoordChange}. Thus, the $a$-function for the zero-temperature metric is 
\begin{equation}\label{eq:HWSM_aFn}
    a(r) = \frac{u(r)^2}{h(r)^{1/2}} \left[\frac{6~h(r)}{2 ~u'(r)~h(r)+u(r)~h'(r)} \right]^3~.
\end{equation}
The derivative of this $a$-function with respect to the energy parameter $\rho$ is given by
\begin{equation}\label{eq:HWSM_RhoDer}
    \frac{da}{d\rho}(r) = \sqrt{u(r)} \frac{da}{dr}
\end{equation}
where (dropping the explicit $r$-dependence)
\begin{equation}\label{eq:HWSM_rDer}
    \frac{da}{dr} = -\frac{108~ u~h^{3/2} }{(u~ h'+2~ u'~h)^4} \left[4~ u~h (u'~ h'+3~ u''~h)+u^2 (6~ h~ h''-5~ h'^2)-8~ u'^2~h^2 \right]~.
\end{equation}

\begin{figure}[h]
    \centering
    \includegraphics[width=0.49\textwidth]{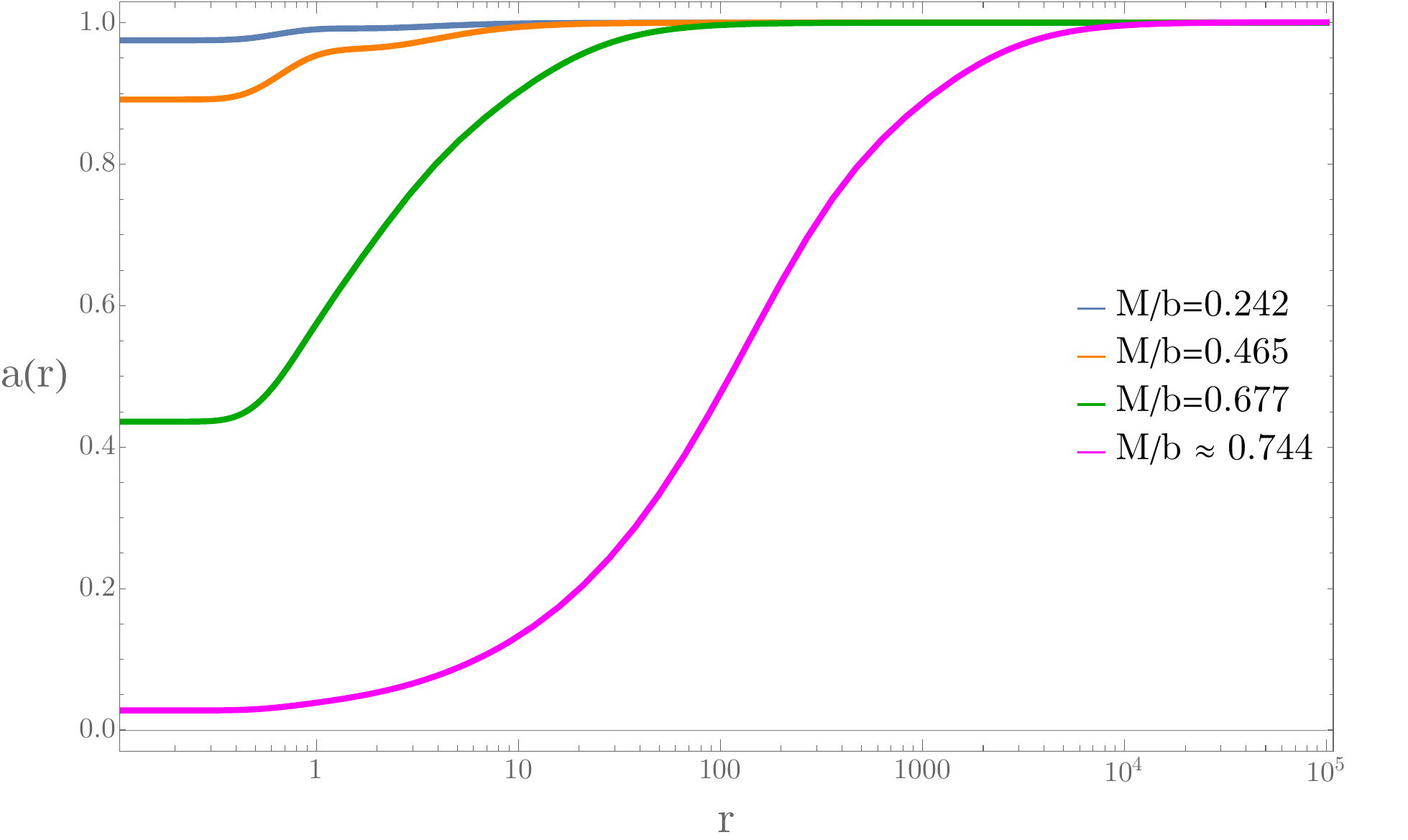}\hfill \includegraphics[width=0.49\textwidth]{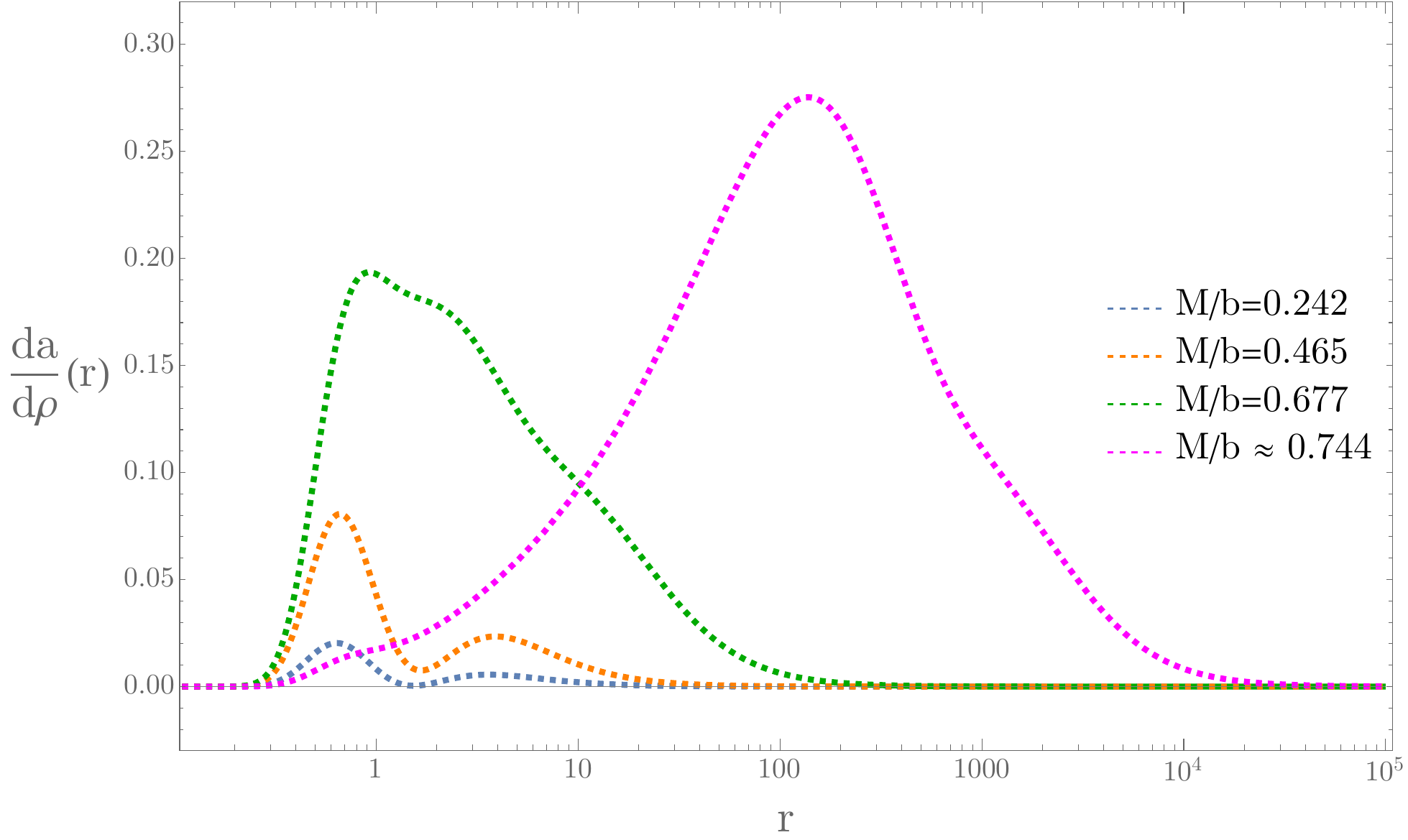}
    \caption{The $a$-function of the HWSM (left) and its derivative $da/d\rho$ (right) for values $M/b<0.744$, corresponding to the topological nontrivial phase. \\}
    \label{fig:HWSMTop}
\end{figure}

\begin{figure}[h]
    \centering
 \includegraphics[width=0.5\textwidth]{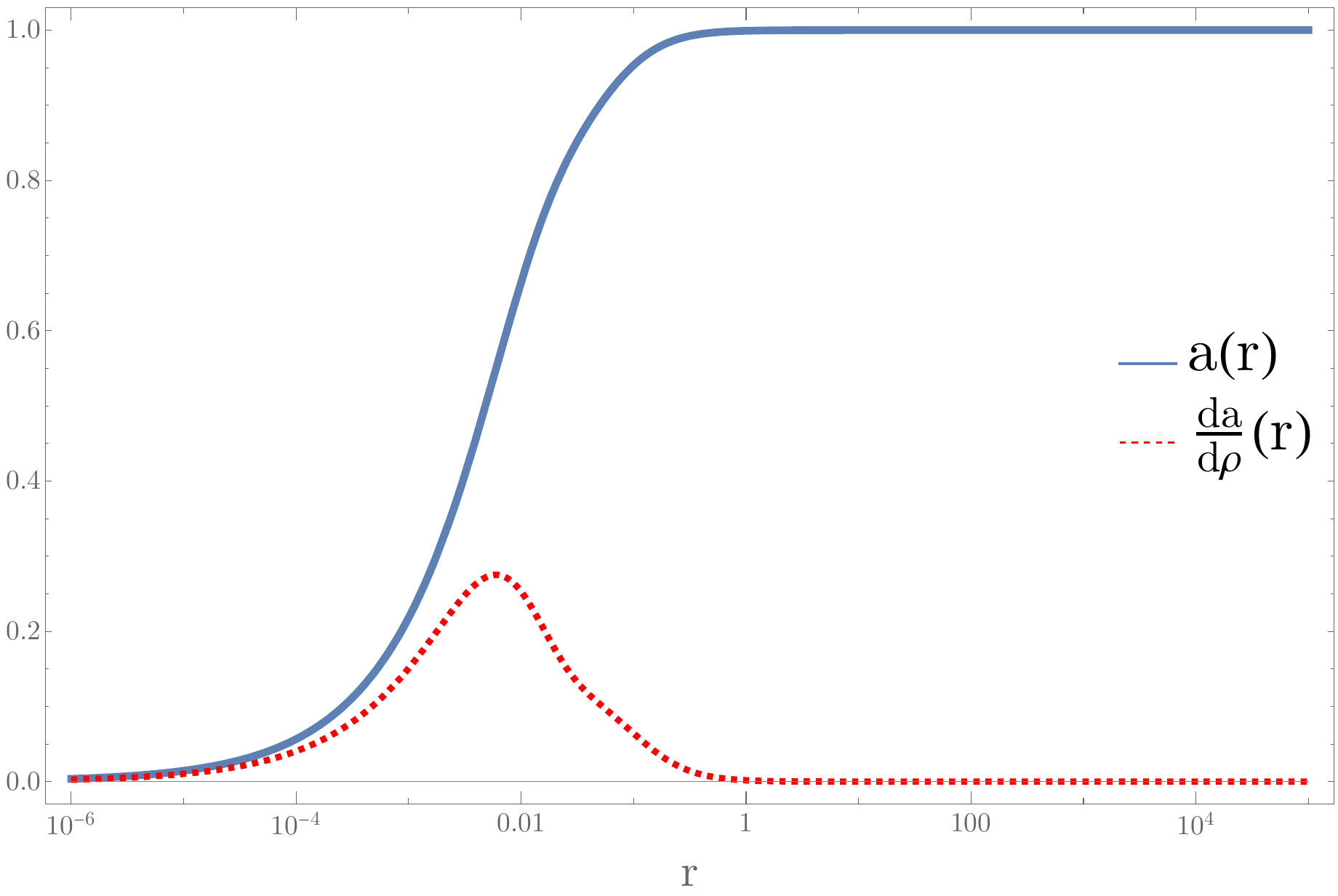}
    \caption{The $a$-function of the HWSM (solid line) and its derivative $da/d\rho$ (dashed line) in the critical point $M/b = 0.744$. \\}
    \label{fig:HWSMCrit}
\end{figure}

\begin{figure}[h]
    \centering
    \includegraphics[width=0.49\textwidth]{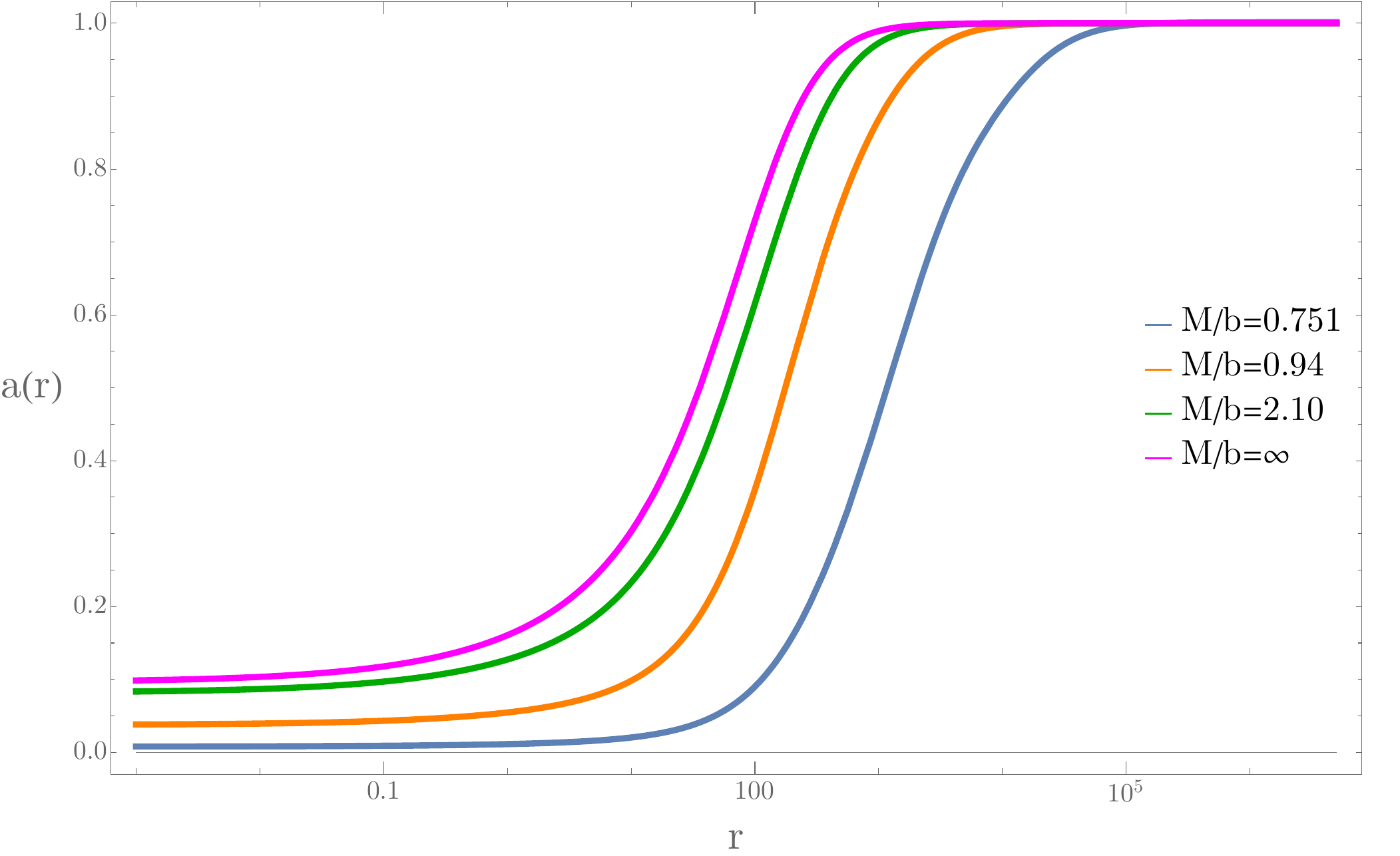}\hfill \includegraphics[width=0.49\textwidth]{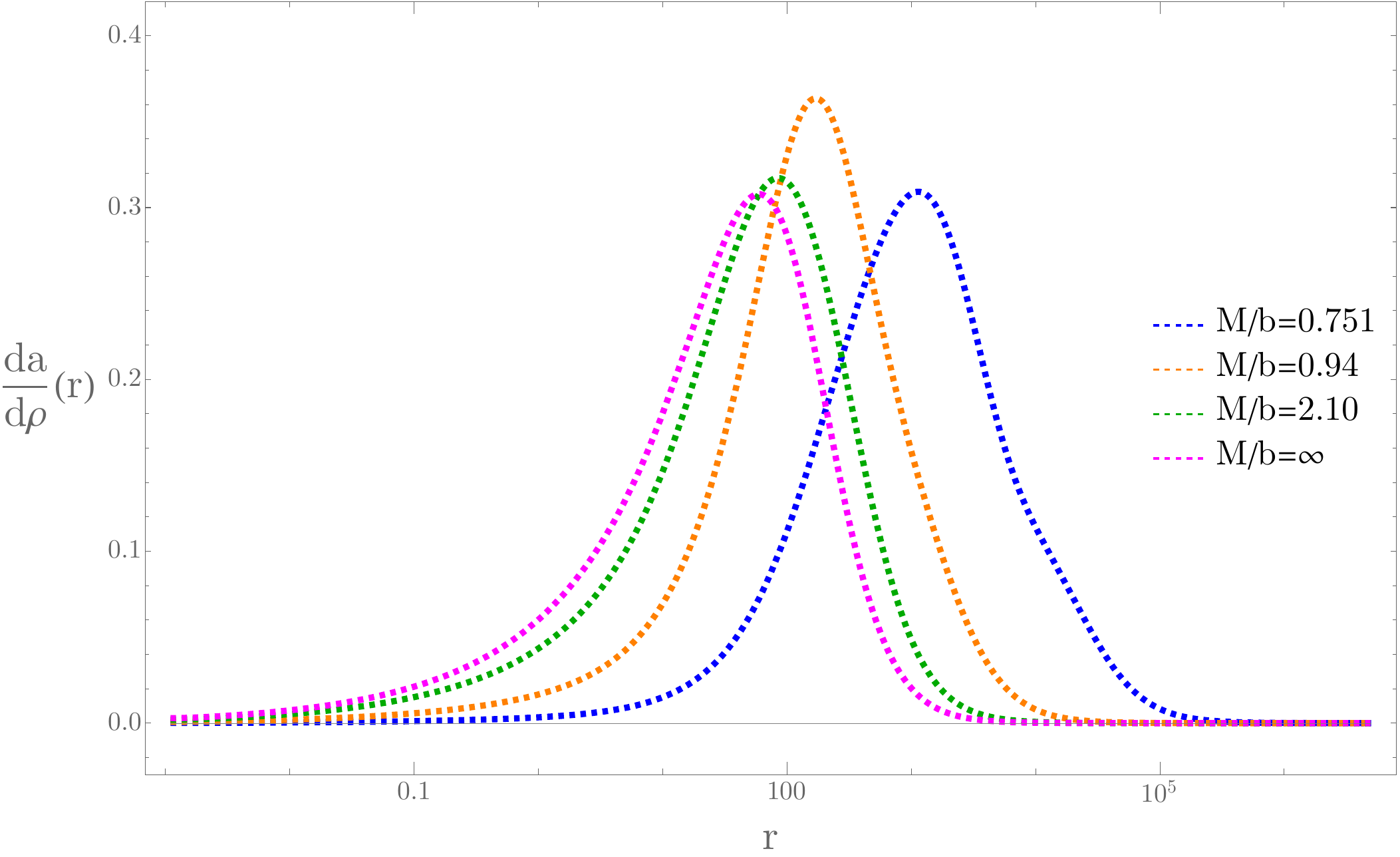}
    \caption{The $a$-function of the HWSM (top) and its derivative $da/d\rho$ (bottom) for values $M/b>0.744$, corresponding to the topological trivial phase. \\}
    \label{fig:HWSMTrivial}
\end{figure}

Here we present the $a$-function and its derivative in terms of $r$ again. The reason for doing so is that the solutions where found numerically in terms of $r$ and because it can be checked by numerical integration of \eqref{eq:HWSM_CoordChange} that the coordinate $r$ corresponds to a rescaling of $\rho$ for all the cases considered. In Figure \ref{fig:HWSMTop} below we show the $a$-function and its derivative for four different values of the parameter $M/b$ that correspond to the topological nontrivial phase. Figure \ref{fig:HWSMCrit} shows the $a$-function and its derivative for the critical point, $M/b=0.744$. Figure \ref{fig:HWSMTrivial} presents the plots for four values $M/b$ that correspond to the trivial phase, including the case of a very large value of the parameter. 
 
\subsection{Analysis of the IR geometries}

In this subsection we will focus on the asymptotic expansion of the geometry towards the IR. There we have analytic expressions that will shed light on the physical interpretation of the $a$-function.

Let us start by siting precisely at the critical point.
As seen in Figure \ref{fig:HWSMCrit}, the $a$-function goes to zero in the IR for those solutions. This can be understood by taking the exact Lifshitz-type solution that describes this case in the IR, with ansatz
\begin{equation}
    u(r)=u_0r^{2\alpha}~,~~~~~~h(r)=h_0 r^{2\beta}~,~~~~~~A_z(r)=A_0r^\gamma~,~~~~~~\phi(r)=\phi_0 r^{\delta}~.
\end{equation}
The equations of motion constraint this ansatz to take the form
\begin{equation}\label{eq:HWSM_CriticalScaling}
    u(r)=u_0r^{2}~,~~~~~~h(r)=h_1 r^{2\beta}~,~~~~~~A_z(r)=A_0r^{2\beta}~,~~~~~~\phi(r)=\phi_0 ~,
\end{equation}
where $h_1=\displaystyle h_0/A_0^2$.

The four parameters $(u_0,~h_1,~\beta,~\phi_0)$ can be written in terms of $(m^2,~q,~\lambda)$ by means of the equations of motion:
\begin{equation}\label{eq:HWSM_ParametersCritical}
    \displaystyle u_0=\frac{2 q^2 \phi_0^2}{3 \beta}~,~~~~~~~~~ h_1=-\frac{q^2}{m^2+\lambda \phi_0^2}~,~~~~~~~~~ \beta = - \frac{2 q^2}{m^2+\lambda \phi_0^2-2q^2}~.
\end{equation}
It is possible to find conditions on the model parameters by plugging \eqref{eq:HWSM_CriticalScaling} and \eqref{eq:HWSM_ParametersCritical} into the $a$-function \eqref{eq:HWSM_aFn} and its derivative \eqref{eq:HWSM_RhoDer}, \eqref{eq:HWSM_rDer} and demanding non-negativity for both. For the $a$-function, we get
\begin{eqnarray} \label{eq:HWSM_aFnCritical}
    a_{crit}(r) = \frac{81 r^{1-\beta} \sqrt{\beta (1-\beta)}}{\sqrt{2}~ (2+\beta)^3~ q^2~ \phi_0^2}~.
\end{eqnarray}
Its derivative is given by
\begin{eqnarray}
    \frac{da_{crit}}{d\rho} (r)=\frac{27 r^{1-\beta} (1-\beta) \sqrt{3(1-\beta)}}{(2+\beta)^3~ q~ \phi_0}~.
\end{eqnarray}

By demanding that the functions are real plus $a_{crit}\geq 0$, $da_{crit}/d\rho \geq 0$ and  $da_{crit}/d\rho \rightarrow 0$ as $r \rightarrow 0$, we obtain the following conditions:
\begin{itemize}
    \item $\beta$, $q$ and $\phi_0$ are real.
    \item $\beta >0 $ and $1-\beta \geq 0$. Thus, $0 < \beta \leq 1$.
    \item $q~\phi_0 > 0$.
\end{itemize}
The fact that $1-\beta>0$ also helps see that the $a$-function goes to zero at $r=0$. For the case $1-\beta = 0$, the $a$-function is zero for all values of $r$. These conditions translate into $(m^2,q,\lambda)$ by means of \eqref{eq:HWSM_ParametersCritical}. We find that $m^2+\lambda \phi_0^2 < 0$. 
We notice that these exact conditions were found in \cite{Copetti_2017} by asking regularity of the IR geometry, reality of the fields and the null energy condition. This is not surprising as they are the basic assumptions behind the positivity and monotonicity of the $a$-function.

We also want to comment on the fact that the $a$-function for the scaling solutions and thus for the critical point vanishes in the deep IR for $\rightarrow 0$. So far such behavior has only been observed for singular metrics. It is indeed well-known that Lifshitz-type geometries suffer from diverging tidal forces as well as other pathologies, although all curvature invariants stay finite. Our case is, however, a bit different in that it is a space-like dimension that has a scaling exponent different from one. Therefore the results about singularities of Lifshitz geometries do not directly carry over. Still, the vanishing of the $a$-function might indicate the presence of some sort of more subtle singularity. The question of if the space like Lifshitz geometries appearing here also present some type of singularities deserves its own treatment but it goes beyond the scope of this article.

Let us go back to the solutions with restored Lorentz invariance in the IR. The asymptotic solutions should read 
\begin{equation}
    u\approx u_0 r^2 ~,~~~~~~h\approx h_0 r^{2}~,~~~~~~A_z\approx A_0r^{2\beta_1}~,~~~~~~\phi\approx\phi_0+\phi_1 r^{\beta_2} ~,
    \label{eq:HWSM_alads}
\end{equation}
This IR  metric corresponds to a Minkowski metric with different light velocity in the $z$ direction with respect to that in the $x-y$ plane. To be precise, the speed of light in the $x-y$ plane coincides with that of the UV fixed point that of course we take to be one. On the other hand, for the $z$ direction we have 
\begin{equation}
    c_z^2=\frac{u_0}{h_0}
\end{equation}
The other important feature of these geometries is the IR effective cosmological constant, that fixes the $AdS_{IR}$ radius. This is the quantity that efectively measures degrees of freedom for relativistic RG flows, and having a relativstic IR we can easily extrapolate
\begin{equation}
    N_{IR}=\frac{1}{u_0^2}
\end{equation}

Now let us plug the ansatz \eqref{eq:HWSM_alads} into the $a$-function we find 
\begin{equation}
    a\to \frac{1}{\sqrt{h_0}u_0}= c_z N_{IR}^{1/4}
\end{equation}
so that even in situations when $N_{UV}=N_{IR}$, the deformation of the light cone accounts for the irreversibility of the RG flow. Notice that if we only account for the IR geometry, a simple re-scaling of the $z$ coordinate allows us to define a perfectly rotation invariant light cone. It is only when we take into account the full RG that such re-scaling is no longer available as it would ruin the UV light cone. 

Finally, notice that at the critical point, the $z$ coordinate has an anisotropic Lifshitz like scaling. This is compatible with taking $c_z=0$ giving an alternative interpretation for the behavior of the $a$-function in the critical domain wall solutions.

\section{Holographic Flat Bands} \label{Section:HFB}


In the context of topological semimetals, another model was recently proposed in \cite{Grandi:2021bsp}. The model is features two Dirac points that approach each other as we increase the temperature. When they collide, they merge into a Berry monopole of charge two, that features a massless quadratic dispersion relation. Following the Weyl semimetal construction, this model is built by reproducing the symmetry breaking pattern of a free model for Dirac fermions with quadratic band touching. Interestingly, the strong interactions inherent to an holographic model lift the inherent state degeneracy of the flattened band by driving the system into a nematic phase characterized by two Dirac cones, which is not possible to describe only with free Fermions. When heating up the system, thermal fluctuations destroy the nematic order parameter. For the Dirac Green function this implies that two poles separated in momentum space collide at zero frequency restoring the rotation symmetry at all scales, and featuring a flattening of the band due to a zero Fermi velocity \cite{Grandi:2021jkj}.

To study irreversibility of the RG flow, we will sit at zero temperature. Hence we will always be in the Dirac phase. This phase breaks rotation invariance spontaneously and only at intermediate scales: full Lorentz invariance is recovered in the deep IR, where we recover a CFT with the same central charge than in the UV. This will be our last example of Boomerang RG flow.


The bulk theory consists of the standard $d+1=4$-dimensional AdS gravitational sector. We also impose that the theory counts with a $SU(2)$ gauge invariance. The minimal action that satisfies such requirements is
 \begin{equation}\label{eq:HFB_BulkAction}
     S = \int d^4x \sqrt{-g}(R-2\Lambda)-\frac{1}{4}\int\Tr(G\wedge *G)~,
 \end{equation}
where  $G=dB-i(q/2) B\wedge B$ is the SU(2) gauge field strength and we set the AdS radius $L=1$. 

To have an asymptotically AdS spacetime in the bulk while leaving the conditions needed to have flat bands, it is necessary to turn on a relevant deformation that breaks the boundary SU(2) group down to U(1). An ansatz for the gauge field $B=B^a \sigma_a/2$ ($a=1,2,3$) is 
\begin{equation}\label{eq:HFB_FieldsAB}
    B=\frac{1}{2} \left[Q_1(r) \sigma_1 + Q_2(r)\sigma_2 \right]dx + \frac{1}{2} \left[Q_1(r) \sigma_2 + Q_2(r)\sigma_1 \right]dy~,~~~~~
\end{equation}
with the boundary condition
\begin{equation}\label{eq:HFB_BdryCondB}
    B \xrightarrow[\tir \rightarrow 0]{} m_* (\sigma_1 dx+\sigma_2 dy)=W~,
\end{equation}
following the convention in \cite{Grandi_2021} where $\tir\geq 0$, with the UV boundary at $\tir=0$.

The ansatz \eqref{eq:HFB_FieldsAB} with the condition \eqref{eq:HFB_BdryCondB} let us break the rotational symmetry in the $xy$ plane as well as the U(2) gauge symmetry, while preserving a combination of the two. It also allows for a nematic phase when $Q_2 \neq 0$, since it breaks the SO(2) rotations down to a discrete $Z_2$.

\begin{figure}[h]
    \centering
    \includegraphics[width=0.46\textwidth]{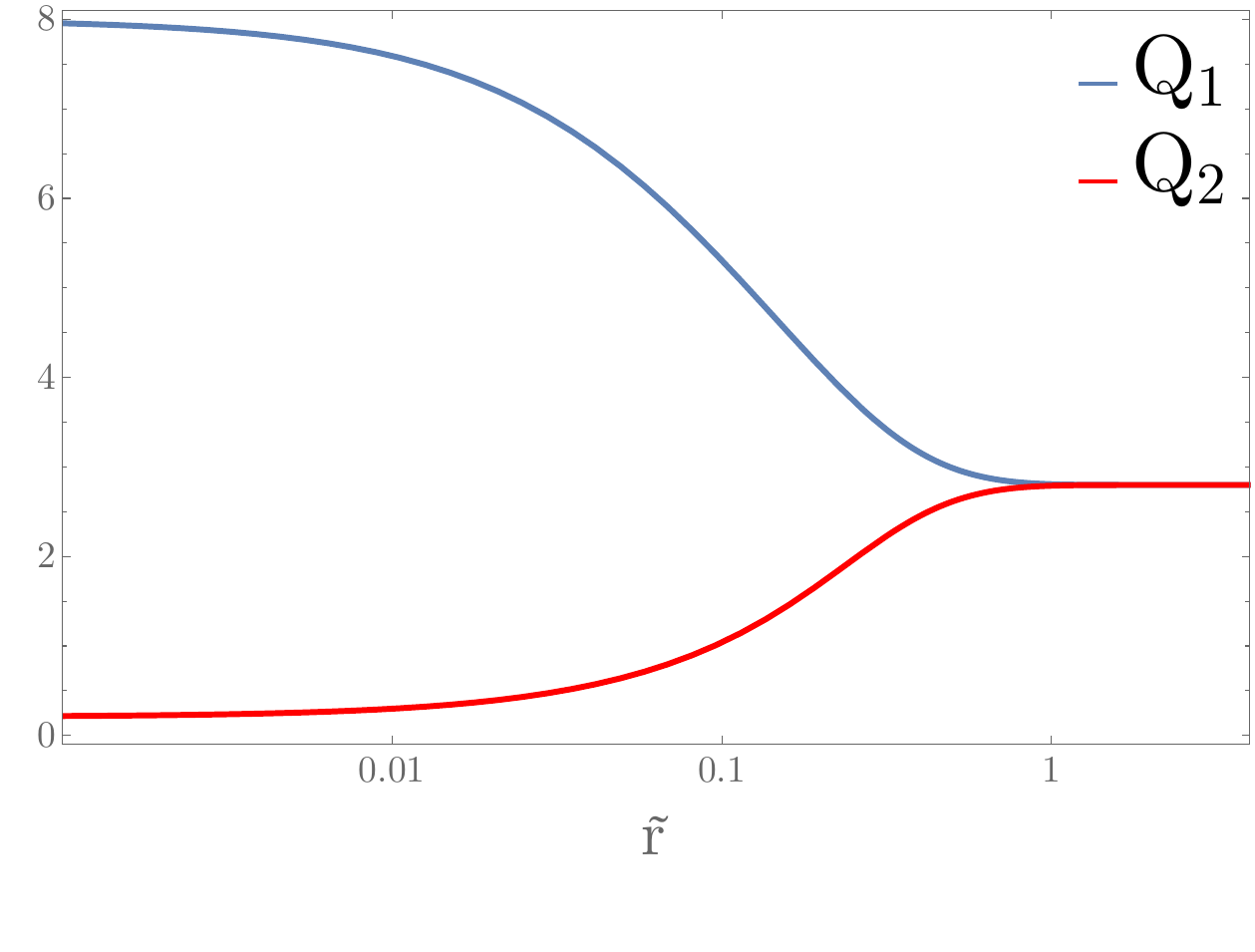}\hfill \includegraphics[width=0.46\textwidth]{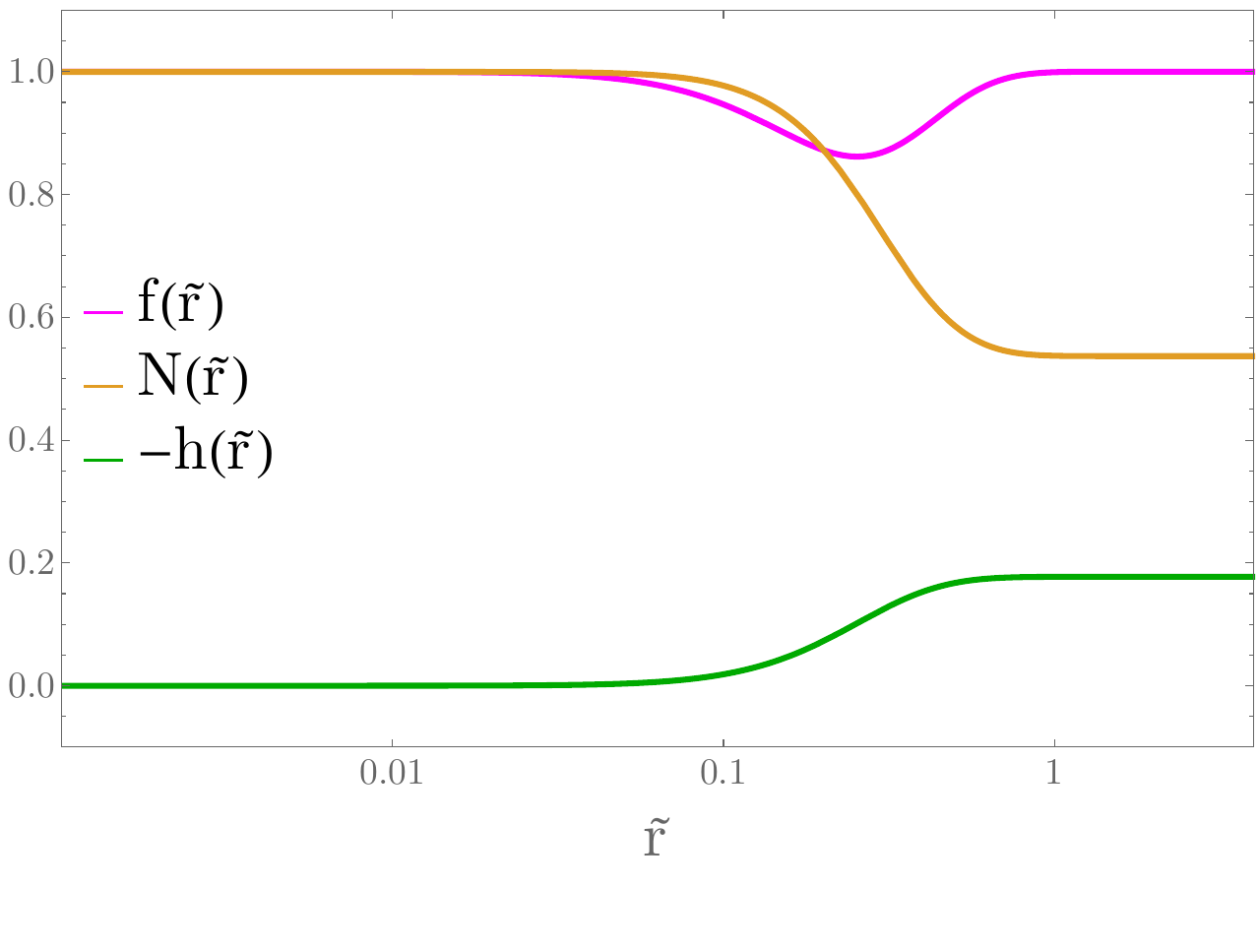}\hfill 
    \caption{Numerical example of a domain wall solution for the metric and Yang-Mill fields. We take the backreaction parameter value to be $q=2$. }
    \label{fig:my_label}
\end{figure}

The bulk metric ansatz is given by
\begin{equation} \label{eq:HFB_BulkMetric}
    \mbox{d}s^2=\frac{1}{\tir^2} \left(-N(\tir) f(\tir) dt^2+\frac{d\tir^2}{f(\tir)} + dx^2 + dy^2 + 2 h(\tir)~ dx~ dy \right)~.
\end{equation}
In order to have and asymptotically AdS spacetime, we impose the conditions
\begin{equation}\label{eq:HFB_BdryCondMetricFuncs}
    f \xrightarrow[\tir \rightarrow 0]{} 1~,~~~~~~~~~~~~~~~N\xrightarrow[\tir \rightarrow 0]{}1~,~~~~~~~~~~~~~~~h \xrightarrow[\tir \rightarrow 0]{} 0~.
\end{equation}
Finally we integrate the equations of motion from an IR geometry at $\tilde r\to \infty$ we will take to be an AdS with the same AdS radius $L=1$. To accomplish this we wil take $Q_1(\infty)=Q_2(\infty)$ in the deep IR that renders a null Yang-Mills curvature hence making pure $AdS$ an obvious solutions. A numerical example is shown in Figure \ref{fig:my_label}.

\subsection{The $a$-function of Holographic Flat Bands}

To obtain the $a$-function for the bulk ansatz \eqref{eq:HFB_BulkMetric}, we first perform a change of coordinates $\tir \rightarrow \rho$ as follows:
\begin{eqnarray} \label{eq:HFB_CoordChange}
    d\rho=-\frac{d\tir}{\tir \sqrt{f(\tir)}}~, &\hspace{0.5in} &e^{2A(\rho)} f(\rho)^2=\frac{N(\tir) f(\tir)}{\tir^2}~, \hspace{0.5in} H(\rho) = h(\tir)~.
\end{eqnarray}
We can rewrite \eqref{eq:HFB_BulkMetric} in the form
\begin{equation}\label{eq:HFB_MetricRewrite}
    ds^2=e^{2A(\rho)} \left[-f( \rho)^2 dt^2+ dx^2 + dy^2 + 2H(\rho)dx~ dy \right]+d\rho^2~.
\end{equation}
Thus, it is straightforward to use \eqref{eq:IdentificationsRNEC} with $d=3$ for this case and obtain the $a$-function
 \begin{equation} \label{eq:HFB_aRho}
 a(\rho) =\displaystyle \frac{1}{\sqrt{1-H(\rho)^2}}~\left[\frac{2 f(\rho)}{2A'(\rho)-\frac{H(\rho) H'(\rho)}{1-H(\rho)^2}} \right]^{2}~.
 \end{equation}
 
\begin{figure}[h]
    \centering
     \includegraphics[width=0.5\textwidth]{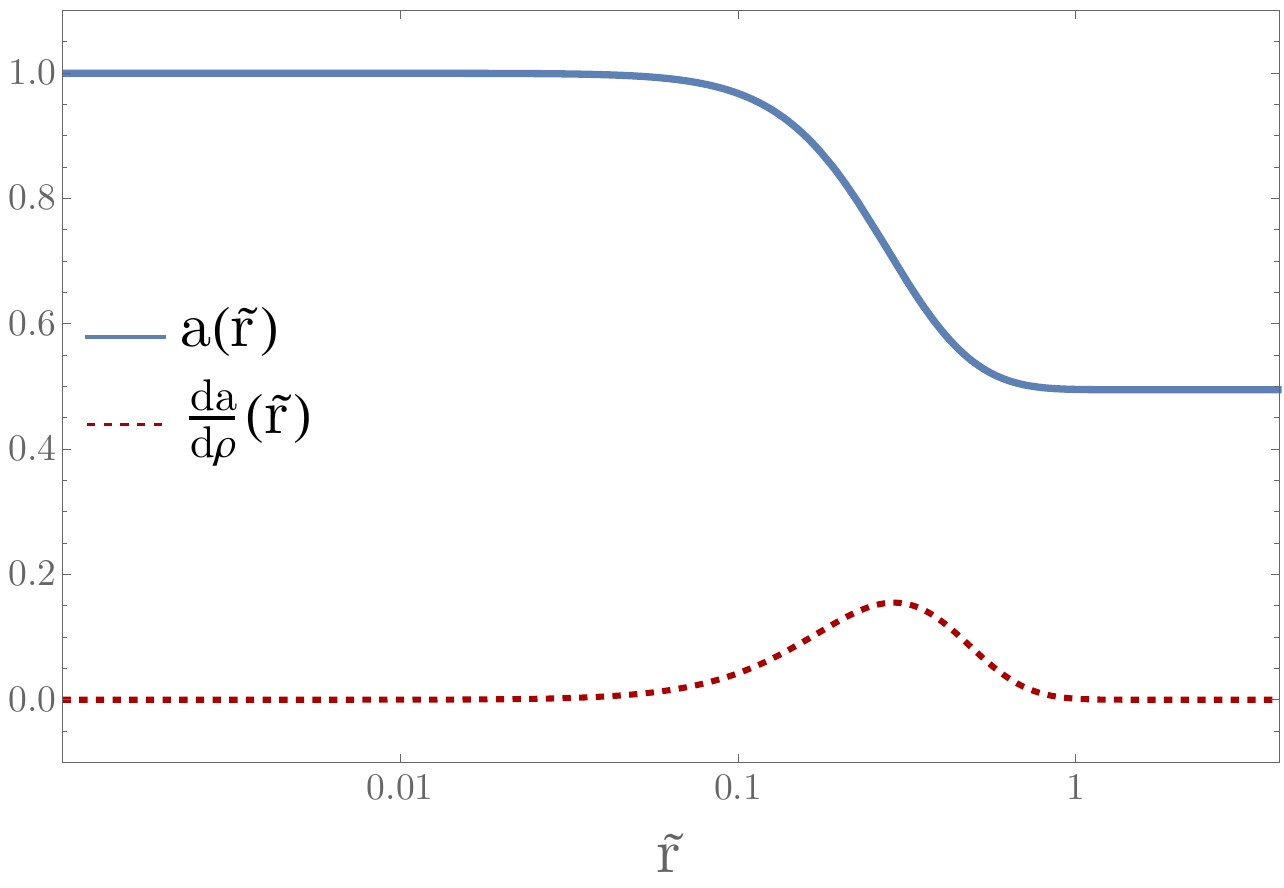}
    \caption{Holographic $a$-function for the solutions displayed in Figure \ref{fig:my_label}}
    \label{fig:my_label2}
\end{figure}

By using the coordinate transformations \eqref{eq:HFB_CoordChange}, we can write the $a$-function of HFB in terms of $\tir$
\begin{equation}\label{eq:HFB_aFn}
    a(\tir) =\frac{4 N(\tir)}{\sqrt{1-h(\tir)^2} \left(2+\displaystyle \frac{\tir~ h(\tir)~ h'(\tir)}{1-h(\tir)^2} \right)^2}~.
\end{equation} 
We calculate the derivative of this $a$-function with respect to the energy scale $\rho$. For this, we use the chain rule and the coordinate transformations \eqref{eq:HFB_CoordChange} to obtain
\begin{equation}\label{eq:HFB_DerARho}
    \frac{da}{d\rho}(\tir) = -\tir \sqrt{f(\tir)} \frac{da}{d\tir}~,
\end{equation}
where the derivative of the $a$-function with respect to $\tir$ is given by (dropping the explicit $\tir$-dependence)
\begin{equation}\label{eq:HFB_DerAr}
    \frac{da}{d\tir} = \frac{4}{\sqrt{1-h^2} \left(2+\displaystyle \frac{\tir~ h~ h'}{1-h^2} \right)^2} \left[N'+\frac{\tir N \left[h'^2 (2+h^2)+2 h h'' (1-h^2) \right]}{(1-h^2)(-2+2 h^2-\tir h h')} \right]\,.
\end{equation}
We explicitily evaluate the $a$-function and its derivative along the RG flow in Figure \ref{fig:my_label2}.

To understand better the physical interpretation of the $a$-function \eqref{eq:HFB_aFn} we can focus on the fixed points of the flow, where we have enhanced symmetries and hence a better chance to make contact with more intuitive concepts. As done in previous examples, we proceed to define the IR speed of light
\begin{equation}
    c_\pm^2 =\frac{N_\infty}{1\pm h_\infty}
\end{equation}
where we have defined $c_\pm$ as the speed of light in the $x_\pm = (x\pm y)/\sqrt{2}$ coordinates for simplicity. Evaluating the $a$-function in the UV one finds that it goes to $1$ close to the UV AdS boundary. In the IR AdS on the other hand, 
\begin{equation}
a\to N_\infty/\sqrt{1-h_\infty^2}= \sqrt{c_+ c_-}
\end{equation}
Hence although the UV and IR AdS radius coincide, giving the same central charges, the $a$-function knows about the RG scale due to the lapse function and shift vector.


\section{Discussion and outlook} \label{Section:Outlook}

In this paper we studied some exotic holographic RG flows featuring a Boomerang behavior triggered by a Poincar\'e breaking relevant deformation. Using the radial null energy condition, we built a monotonic $a$-function. In the UV, this $a$-function simply measures the effective AdS radius. On the other hand, the behavior of the $a$-function in the IR depends both on the effective AdS radius and the speed of light, making it a sensible $a$-function.

Let us briefly comment on some open questions we would like to address in the future

\begin{itemize}
    
    \item The holographic Weyl semimetal has a free model counterpart, featuring similar physics. This model is a perfect candidate to gain a more field theoretical insight in the problem of non-relativistic RG flows.

    \item The question of irreversibility in the presence of Lorentz breaking relevant deformations still escapes a general treatment. Our findings suggest that Boomerang RG flows, though looking exotic at first sight, might be a good playground to make some first steps in this complicated arena.

    \item We found that the $a$-function goes to zero for our solutions with a Lifshitz like scaling. Traditional Lifshitz geometries are known to suffer from mild divergencies. It would be interesting to find out if such divergencies are present in this modified Lifshitz geometries. If so, it would be good to understand if the peculiar behavior of the $a$-function in these regimes is somehow related to that fact.

    \item These $a$-functions were recently motivated by the study of the so called trans-IR RG flows \cite{Caceres:2022smh,Caceres:2022hei,Caceres:2023zhl}. In this letter we analyze zero-temperature models only, but it would be interesting to extend our exploration beyond black hole horizons.

    \item AdS/CMT provides of a plethora of geometries where general relativity concepts such as the NEC can be tested, and its consequences to their holographic dual conjectured and sometimes elucidated. In this work we studied geometries with broken rotation invariance in the boundary coordinates. This translates into different shift vectors in the corresponding coordinates of the metric. An interesting extension of our work would be to include $f(r)dt dx$ type of terms into the metric. These terms are relevant in the context of holographic superfluids with a superflow \cite{Arean:2011gz}. This kind of terms also appears in the context of Dyonic black holes with a Chern-Simons term \cite{DHoker:2009ixq}.

\end{itemize}

\acknowledgments

The authors would like to thank Lorenzo Di Pietro, Carlos P\'erez-Pardavila, Sanjit Shashi and Aaron Zimmerman for useful discussions. The work of EC is supported by the National
Science Foundation under Grant Number PHY-2112725.  The work of RCV is supported by the Robert N. Little Fellowship.  K.L. is supported through the grants CEX2020-001007-S and PID2021-
123017NB-100, PID2021-127726NB-I00 funded by MCIN/AEI/10.13039/501100011033
and by ERDF “A way of making Europe". ISL is a CONICET and ICTP Associate (2023-2028) fellow.
EC thanks the Instituto de F\'isica Te\'orica (IFT) at UAM, Madrid,
for hospitality during the initial stages of this work. ISL thanks to ICTP, IFT and Católica Chile U. for hospitality.


\bibliographystyle{JHEP}
\bibliography{refCalc2}
 \end{document}